\documentclass[11pt]{article}

\usepackage{times}
\usepackage{setspace}
\usepackage{titletoc}
\usepackage{color}
\usepackage{natbib}
\usepackage[subrefformat=parens,labelformat=parens]{subfig}
\usepackage{amssymb, amsmath, amsthm, amsfonts, array, multirow}
\usepackage{graphicx, float}
\usepackage[skip=10pt, font=small, labelfont=bf]{caption}
\usepackage[toc,page]{appendix}
\usepackage{url}
\usepackage[margin=1in]{geometry}
\usepackage{titling}
\usepackage[dvipsnames]{xcolor}
\usepackage[inline]{enumitem}
\usepackage{adjustbox}
\usepackage{bm}
\usepackage[colorlinks=true,allcolors=blue]{hyperref}
\usepackage[capitalize,noabbrev]{cleveref}
\crefformat{equation}{(#2#1#3)}

\newtheorem{assumption}{Assumption}

\DeclareMathOperator{\logit}{logit}

\newcommand\blfootnote[1]{%
  \begingroup
  \renewcommand\thefootnote{}\footnote{#1}%
  \addtocounter{footnote}{-1}%
  \endgroup
}

\makeatletter
\newcommand{\printfnsymbol}[1]{%
  \textsuperscript{\@fnsymbol{#1}}%
}
\makeatother

\title{\Large Evaluating Federal Policies Using Bayesian Time Series Models: Estimating the Causal Impact of the Hospital Readmissions Reduction Program}

\author{Georgia Papadogeorgou\thanks{Equally contributing authors} \thanks{University of Florida, Department of Statistics, Gainesville, FL, USA. 
ORCID 0000-0002-1982-2245. 
Corresponding Author. gpapadogeorgou@ufl.edu.}, \ \ 
Fiammetta Menchetti\printfnsymbol{1}\thanks{University of Florence, DiSIA and Florence Center for Data Science, Florence, Italy. ORCID 0000-0003-3463-0206.
}, \ \ 
Christine Choirat\thanks{Swiss Data Science Center, ETH Zürich and EPFL, Lausanne \& Institute of Global Health, Faculty of medicine, University of Geneva, Switzerland.
}, \\
Jason H. Wasfy\thanks{
Cardiology Division, Department of Medicine, Massachusetts General Hospital, Harvard Medical School, Boston, MA, USA}, \ \ 
Corwin M. Zigler\thanks{University of Texas at Austin, Department of Statistics and Data Science, Department of Women's Health, Austin, TX, USA}, \ \ 
Fabrizia Mealli\thanks{University of Florence, DiSIA and Florence Center for Data Science, and European University Institute, Department of Economics, Florence, Italy.  ORCID 0000-0002-1076-9249} 
\\[20pt]}

\date{\vspace{-2cm}}

\allowdisplaybreaks

\begin{document}



\maketitle

\begin{abstract}
 \blfootnote{The authors would like to thank Francesca Dominici, Alessandra Mattei, and Joseph Antonelli for their constructive comments.}
Researchers are often faced with evaluating the effect of a policy or program that was simultaneously initiated across an entire population of units at a single point in time, and its effects over the targeted population can manifest at any time period afterwards. In the presence of data measured over time, Bayesian time series models have been used to impute what would have happened after the policy was initiated, had the policy not taken place, in order to estimate causal effects. However, the considerations regarding the definition of the target estimands, the underlying assumptions, the plausibility of such assumptions, and the choice of an appropriate model have not been thoroughly investigated. In this paper, we establish useful estimands for the evaluation of large-scale policies.
We discuss that imputation of missing potential outcomes relies on an assumption which, even though untestable, can be partially evaluated using observed data.  We illustrate an approach to evaluate this key causal assumption and facilitate model elicitation based on data from the time interval before policy initiation and using classic statistical techniques.
As an illustration, we study the Hospital Readmissions Reduction Program (HRRP), a US federal intervention aiming to improve health outcomes for patients with pneumonia, acute myocardial infraction, or congestive failure admitted to a hospital. We evaluate the effect of the HRRP on population mortality among the elderly across the US and in four geographic subregions, and at different time windows. We find that the HRRP increased mortality from pneumonia and acute myocardial infraction across at least one geographical region and time horizon, and is likely to have had a detrimental effect on public health.
\end{abstract}

\textbf{keywords:} Bayesian methods, causal inference, hospital readmissions reduction program, mortality, policy evaluation, time-series

\section{Introduction}

Researchers and policy makers are often interested in understanding the effect of new policies or programs. The target policy is often rolled out at once over the whole region of interest, it alters the operation of a group of organizations, which might affect the region's population in a number of ways. To evaluate the policy, available data include a time period before policy rollout (the ``pre-intervention'' period), and a time period after policy initiation (the ``post-intervention'' period). Considering that the policy rollout is experienced by everyone in the region, policy evaluation should also target the policy's impact on the whole population. Therefore, we are faced with estimating the causal effect of a policy over a single unit measured over time, including a pre- and a post-intervention time period.

In the presence of pre- and post-intervention data, popular methods include the difference-in-difference \citep{Athey2006} and synthetic control~\citep{Doudchenko2017} approaches. However, these approaches require the presence of control units during both time periods, whereas in our case, everyone is exposed to the policy once it is initiated. 
Other measured time series for the same population have been previously used as assumed control units in synthetic control methodology \citep{gaughan2019paying}. 
Interrupted (or quasi-experimental) time-series analysis has been alternatively used for estimating causal effects from a single time series \citep{mcdowall2019interrupted}, but it requires that we correctly specify the outcome model in both the pre- and in the post-intervention time period, an assumption that is rather restrictive. \cite{bojinov2019time} introduced causal estimands and estimation based on a single time series, but their methodology requires that the treatment can be applied and taken away at any time point, which is not true for policy settings where once the policy is initiated it remains active throughout.

In this manuscript, we suggest using the whole region where the intervention took place as a single system (unit) for policy evaluation. Under the potential outcome framework for causal inference, we define causal estimands that describe the effect of the policy on the whole system, and introduce the crucial causal assumptions for estimating these quantities using observed data. For estimation, we employ Bayesian time series modeling to impute what would have happened in the post-intervention time period, had the intervention not taken place, and we combine these imputations with observed outcomes to unbiasedly estimate causal effects, similarly to other approaches in the literature~\citep{Brodersen2015, Miratrix2019simulating, antonelli2020heterogeneous}. We illustrate how model choice and the plausibility of the causal assumptions can be partially evaluated using data from the pre-intervention time period and classic statistical techniques, and 
using time series data that are expected to be unaffected by the intervention. 
The framework presented here provides guidance for the evaluation of programs and policies of various sorts, such as educational policies, social programs, or environmental policies that are adopted by schools, neighborhoods, and polluting sources, respectively, and affect the outcomes of the region's population \citep{zigler2021bipartite}.

This work is motivated by the evaluation of the Hospital Readmissions Reduction Program (HRRP), part of the Affordable Care Act which passed in 2010. The HRRP created the prospect of financial penalties that would start being applied in 2012 based on risk-standardized readmission rates for Medicare fee-for-service patients within 30 days of discharge after index hospitalization for acute myocardial infarction (AMI), congestive heart failure (CHF), or pneumonia. The HRRP reflected intent to incentivize providers and hospitals to improve the coordination and quality of care for patients.
Even though the law was associated with substantial declines in risk-standardized readmissions~\citep{Zuckerman2016, Desai2016, Wasfy2017,MedPAC2018}, whether the law was associated with increased mortality is very controversial~\citep{Joynt2011,Joynt2012,Fonarow2017} with studies returning drastically different results~\citep{Gupta2017, Dharmarajan2017, Wadhera2018, Khera2018, Huckfeldt2019, Sandhu2019}. 
Improving our understanding of the causal effects of the HRRP is critical, because this type of information will inform whether similar policies should be implemented or not in the future. Although most of the aforementioned papers adopt methodologies that are often employed in the causal inference literature (e.g., interrupted time series), the underlying assumptions are not explicit and the estimands are not clearly defined. This is necessary to give a reader an understanding of the effect one is aiming to estimate and under which assumptions it can be interpreted as “causal.”
Within our framework, we define and estimate the causal effect of the HRRP on mortality from AMI, CHF or pneumonia among individuals that were 65 years old or older across the United States and in four subregions, and at three time windows after the passage of the law. We find that the HRRP caused an increase in mortality from AMI across the US and from pneumonia in some geographical subregions.






\section{Methods}

\subsection{Evaluating a policy at the same level of its scope}

For ease of illustration, we focus on the HRRP throughout as a policy which affects hospital operations in an effort to improve public health.
To evaluate such a policy, the prevailing analysis approach compares the outcomes of hospitals that were affected by the policy at varying degrees. However, such an analysis is faced with many challenges. Firstly, hospitals that adopt the policy at different degrees can have systematic differences due to both measured and unmeasured factors which might confound the estimation of causal effects on hospital-level outcomes. Second, the extent to which a hospital adopts the policy might affect the outcome of other hospitals, giving rise to interference, which complicates how causal effects are both defined and estimated. That can happen if, for example, hospitals with high financial penalties under the HRRP avoid admitting high-risk patients, who are afterwards admitted at a different hospital~\citep{casalino2007will}. Lastly, even if effects of the intervention on hospital-level outcomes were properly defined and estimated, they would represent the effects of policy adoption of {\it different degrees} rather than the total effect of the policy, and conclusions on population-level effects based on the analysis of hospital-level outcomes might suffer from ecological bias.

Understanding the effects of policy adoption at different degrees can be essential for uncovering key features of the policy, but the challenges in the analysis of hospital-level data presented above have not been definitively resolved and interpretation of estimated quantities remains complicated. Here, we advocate a complementary approach that bypasses those specific issues in focusing on different, but related, causal estimands. We suggest that evaluation of large-scale policies is performed at the level of the whole population, by viewing the whole region as {\it one system} measured over time, and 
considering causal estimands that describe the change in the outcome over the whole population in the presence and in the absence of the intervention.
As we discuss below, this entails notions of confounding that do not necessarily require adjustment for measured differences. It also bypasses the issue of interference, since the whole population composes a single unit that is treated by the policy, and we observe the outcome at an aggregate level irrespective of which hospital provided care to whom. Finally, and importantly for policy evaluation, the framework below allows us to quantify the effect of the policy directly on the population under study, and therefore does not suffer from ecological biases. 
We find that providing a complimentary way to evaluate large-scale policies by shifting the goalpost from hospital-level analyses to focusing on estimands over the whole population is a key contribution of our work.


\subsection{Causal estimands in single time series with one interventional time point}
\label{sec:potential_outcomes}

Since the policy could have been passed or not, we formulate our causal estimands using two potential time series: one representing the outcome had the policy taken place, which is observed, and one representing the outcome had the policy not taken place, which is not observed and has to be estimated. 

Let $t \in \mathcal{T} = \{1, 2, \dots, T\}$ represent $T$ time periods of observed data, and $t^*\in \mathcal{T}$ denote the time period during which the policy was actually initiated. Let $w_t = 1$ denote that the policy is active at time period $t$, and $w_t = 0$ that it is not. Then,
\begin{equation}
\overline w_{1T} = (w_1 = 0, w_2 = 0, \dots, w_{t^* - 1} = 0, w_{t^*} = 1, w_{t^*+1} =1, \dots, w_T = 1) 
\label{eq:trt_presence}
\end{equation}
represents the observed scenario where the policy is initiated at time $t^*$ and remains active thereafter,
and
\begin{equation}
\overline w_{0T} = (w_1 = 0, w_2 = 0, \dots, w_{t^* - 1} = 0, w_{t^*} = 0, w_{t^*+1} =0, \dots, w_T = 0) 
\label{eq:trt_absence}
\end{equation}
represents the counterfactual scenario where the policy is never initiated.
We postulate the existence of potential outcomes $Y^{jt} = Y^t\big(\overline w_{jT}\big)$ for $j \in \{0, 1\}$, which represent the population outcome that would have been observed at time period $t$ had the policy never been initiated, or had it been initiated at $t^*$, respectively.
Then, $Y^{1t} - Y^{0t}$ represents the effect of initiating the policy at time $t^*$ against not initiating it at all on the outcome at time period $t$, similarly to the causal estimands in the synthetic control literature \citep[e.g.,][]{Doudchenko2017}. We focus at time periods after policy initiation, $t \geq t^*$. Then, the causal effect of the policy $k$ time periods after initiation is defined as
\begin{equation}
\Delta_k = Y^{1(t^* + k)} -  Y^{0(t^* + k)}, \ k \geq 0,
\label{eq:effect_at_t}
\end{equation}
and the cumulative effect during the first $K$ time periods is defined as
\begin{equation}
C\Delta_K =
\sum_{k = 0}^{K-1} \Delta_k =
\sum_{k = 0}^{K-1} \Big[ Y^{1(t^* + k)} -  Y^{0(t^* + k)} \Big].
\label{eq:cum_effect}
\end{equation}
In our study, the potential outcomes $Y^{jt}$ represent the total number of deaths across the US had the HRRP passed or not, and the cumulative effect in \cref{eq:cum_effect} represents the number of deaths caused or prevented by the policy within $K$ time periods from its initiation.


Defining potential outcomes at the level of the whole region obviates the need to consider interference explicitly. In contrast, hospital-level potential outcomes should be indexed by the treatments of all hospitals which interfere with it, which would complicate the definition of subsequent hospital-level estimands.
As also highlighted in \citet{Imbens:Rubin:2015}, a common strategy to deal with interference between statistical units is changing the level of analysis.
Therefore, considering the whole region under study as {\it one unit} for which the policy would either take place or not allows us to bypass this issue.

\subsection{The causal assumptions}
\label{subsec:the_assumptions}

Since the policy was indeed initiated at time period $t^*$, the observed treatment is equal to the treatment path $\overline w_{1T}$, and the observed outcome $Y_t$ is equal to the corresponding potential outcome, $Y_t = Y^{1t}$. This implies that all potential outcomes corresponding to the treatment path $\overline w_{1T}$ are observed, as illustrated in the top row of \cref{tab:pot_out}. Therefore, all potential outcomes $Y^{1t}$ in the estimands (\ref{eq:effect_at_t}), (\ref{eq:cum_effect}) are observed, and we are faced with estimating the potential outcomes $Y^{0t}$ in the post-intervention period, $t \geq t^*$. To do so, we make the following assumptions. The interpretation of these assumptions as well as how we can assess their plausibility is discussed in more detail in Sections \ref{subsec:confounding} and \ref{subsec:plausibility_assumpt}, and within the context of our study in \cref{subsec:model_eval}.

\begin{assumption}
The policy does not alter potential outcomes before its initiation, $Y^{0t} = Y^{1t}$ for $t < t^*$.
\label{ass:no_anticipation}
\end{assumption}

\cref{ass:no_anticipation} states that the intervention could not have affected the outcome before it was actually initiated. Therefore, potential outcomes in the pre-intervention time period are the same, irrespective of whether the policy is initiated or not in the future. This assumption establishes that the observed outcomes in the pre-intervention period are equal to the potential outcomes in the absence of the intervention, $Y^{0t} = Y_t$ for $t < t^*$.
The next assumption allows us to use the pre-intervention outcomes to learn the outcome process in the absence of the treatment, potentially conditional on covariates, and is discussed in more detail in \cref{subsec:confounding}.


\begin{table}[!t]
    \centering
\caption{Potential outcomes in the pre- and post-intervention time period, and the causal assumptions.}
\label{tab:pot_out}
    \begin{tabular}{c|cc}
     & pre-intervention period $t < t^*$ & post-intervention period $t \geq t^*$ \\ \hline
$Y^{1t}$ & observed    & observed \\
$Y^{0t}$ & $=$ observed (\cref{ass:no_anticipation}) &  imputed based on \cref{ass:model}
    \end{tabular}
\end{table}

\begin{assumption} 
There exist measured covariates $X_t$ which are unaltered by the intervention, and conditional on which the outcome process is stationary with some lag $M \geq 0$, 
i.e. 
if $X^{jt} = X^t(\overline w_{jT})$ are the potential covariate values under treatment path $\overline w_{jT}$ for $j = 0, 1$, we have that $X^{0t} = X^{1t}$, 
and there exists $M \geq 0$ such that, for all $t \in \mathcal{T}$,
\vspace{-0.5\baselineskip}
\begin{align*}
& P(Y^{0t}|Y^{0(t-1)}, \dots, Y^{0(t - M)}, X^{0t}, \dots, X^{0(t - M + 1)}) \\
& \hspace{20pt} = P(Y^{0(t+1)}|Y^{0t}, \dots, Y^{0 (t-M+1)}, X^{0(t + 1)}, \dots, X^{0(t  - M + 2)}).
\end{align*}
\label{ass:model}
\end{assumption}

\vspace{-1.6\baselineskip}
\cref{ass:model} is the basis for imputing the missing potential outcomes in the absence of the policy. 
For illustration, assume that it holds for $M = 1$ and in the absence of covariates. Then, the conditional outcome distribution is the same for all time points in the pre-intervention period, i.e. for $t_1, t_2 < t^*$ it holds that $P(Y^{0t_1}|Y^{0(t_1-1)}) = P(Y^{0t_2}|Y^{0(t_2-1)})$. Since the potential outcomes $Y^{0t}$ in the pre-intervention period are observed from \cref{ass:no_anticipation}, this conditional distribution can be estimated from the data. Then, setting $t = t^*$ in \cref{ass:model}, we have $P(Y^{0t^*}|Y^{0(t^*-1)}) = P(Y^{0(t^*-1)}|Y^{0(t^*-2)})$ which allows us to impute $Y^{0t^*}$. Once $Y^{0t^*}$ is imputed based on this distribution, the same statement for $t = t^* + 1$ will be used to impute $Y^{0(t^*+1)}$ and so forth. Therefore, \cref{ass:model} enables us to learn the potential outcome process using the observed outcomes in the pre-intervention period, and use it to predict the missing potential outcomes in the post-intervention period had the policy not taken place.

\subsection{Confounding with a single unit measured over time}
\label{subsec:confounding}

Here, we start by discussing the notion of confounding in the setting of a single unit followed over time, and then we relate it to our \cref{ass:model}. An analogous discussion in the context of multiple units measured over time is given in \citet{antonelli2020heterogeneous}.

In most settings in causal inference, measured covariates are assumed to satisfy a no-unmeasured confounding assumption in that they represent all the ways in which treated and control units differ, and that, within cells defined by these covariates, treatment assignment is independent of any potential outcome. In our setting, the whole system is viewed as a {\it single} unit subject to the intervention for which we have access to a time period before the intervention when the unit is untreated, and a time period after the intervention when the unit is treated. Construing the different time periods as the most primitive observational units, confounding would represent systematic differences in the pre- and post-intervention time periods.
Then, the primary threat to validity with estimating the missing potential outcome $Y^{0t}$ in the post-intervention time period is parsing changes due to the policy from other temporal changes that would have occurred regardless.  In the HRRP, this could correspond to secular trends in population lifestyle factors or improvements in medical care technology that are not related to the policy, but may have changed mortality outcomes over time.
To estimate the causal effect by comparing the outcomes in the pre- and post-intervention time period, we would have to account for all such variables through matching, sub-classification, regression, or other methods.

Rather than finding pre- and post intervention time points with similar values of covariates (which would not generally be available) to account for factors that vary coincidentally with the policy and outcomes, \cref{ass:model} focuses on the potential outcomes in the absence of the intervention and states that secular and periodic trends in the pre-intervention outcome time-series would have persisted in the post-intervention period, had the intervention not taken place. Therefore, this assumption could be satisfied if temporal trends in the pre-intervention time period are properly accommodated using flexible functions of time or covariates.
Flexible functions of time can be used to account for the overall decreasing trends in mortality during the pre-intervention period reported in \cref{fig:ts_regional},
without requiring to have access to all variables that might drive these trends.
In fact, flexible temporal trends
can adequately capture covariate time series that predict the outcome and have long term temporal trends, such as population comorbidities which have overall persistent trends during our study period.
Therefore, the assumption might hold without any covariates, in which case covariates are not needed for unbiased estimation of causal effects.
However, researchers have to be mindful to include covariates that predict the outcome time series and have a temporal pattern that changes post-intervention, since these covariates will not be captured by a flexible temporal trend.
Importantly, the covariates that are used in the outcome model should {\it not} be influenced by the treatment themselves.
For example, in a hypothetical study which evaluates the effect of a transportation policy to reduce air pollution, the number of circulating cars should not be used as a covariate, as it is likely to be affected by the intervention.

As a result, the potential drawback of the proposed approach lies in that it cannot capture changes in the temporal trends of the counterfactual time series after the intervention which occur due to unpredictable changes in missing predictors, or other co-occurring causes. Causal inference methodologies for panel data with control time series partially alleviate this issue. For example, synthetic control methodology could accommodate unpredictable changes in the counterfactual time series as long as these changes in the temporal and secular trends manifest similarly in the treated and control units, and regression approaches with treated and control units can improve effect estimation in the presence of co-occurring policies, as long as these policies are not enacted in near succession \citep{Griffin:2022}.
Similarly to our approach, synthetic control and difference-in-differences methodologies are also based on alternatives to the classic no-unmeasured confounding assumption, though they rely on control time series for effect estimation. Synthetic control methodology requires that the weights learnt from data in the pre-intervention period are stable into the post-intervention period, and difference-in-difference methodology is based on the famous parallel trends assumption which, even though can hold conditional on covariates, does not require that all confounders are measured \citep{Angrist:Pischke:2008}.

\subsection{Assessing the plausibility of the causal assumptions}
\label{subsec:plausibility_assumpt}

\cref{subsec:the_assumptions} formalizes the assumptions necessary for unbiased estimation of causal effects. Even though Assumptions \ref{ass:no_anticipation} and \ref{ass:model} cannot be formally tested, they can be qualitatively and quantitatively evaluated based on the observed data.
We discuss how to assess the plausibility of the causal assumptions.

\cref{ass:no_anticipation} describes that outcomes before the interventional time period could not have been altered in {\it anticipation} of the intervention.
Therefore, to satisfy this assumption, the interventional time point $t^*$ should be set as the first time period during which the policy might have altered the outcome under study. That might be at the time of policy implementation, or even earlier when the policy is passed or first announced.
Specifying $t^*$ earlier than necessary will not bias effect estimation; instead, estimation will not be necessarily unbiased if $t^*$ is not set early enough \citep[see][for an approach that identifies the time period when the treatment effect materializes]{cruz2017robust}. 
An approach to quantitatively assess the plausibility of \cref{ass:no_anticipation} is to use the proposed method with an earlier intervention time period, and ensure that the estimated effect during the pre-intervention period is null.

Secondly, Assumption~\ref{ass:model} allows for the use of covariates that are unaltered by the policy.
To check the plausibility that our covariates are not affected by the policy we could, in principle, test whether the intervention impacted the considered covariates by replacing the outcome under study with each covariate and repeating the policy evaluation. In practice, this step can be avoided in many instances. For example, exogenous covariates can often be used straightforwardly. In the study of the HRRP, such covariates could represent weather information which is clearly not affected by the policy. 
After a set of covariates that are not affected by the intervention has been chosen, the pre-intervention outcome data can be used to evaluate \cref{ass:model} using classic statistical tools. 
We suggest using posterior predictive checks \citep{Gelman:Meng:Stern:1996, Gelman:Carlin:Stern:Dunson:Vehtari:Rubin:2013} to do so. Posterior predictive checks involve generating replicated data by drawing from the posterior predictive distribution and comparing these draws to the observed data using both numerical and graphical checks. A simple graphical check plots the mean of the posterior predictive distribution against the observed data. After obtaining the posterior predictive distribution of a given test quantity (e.g., mean, median, maximum), a numerical check could be based on the so-called Bayesian p-value defined as the proportion of times that the observed test quantity exceeds the one of the replicated data. In general, an extreme p-value (very close to 0 or 1) indicates that the feature of the data captured by the test quantity is inconsistent with the assumed model.
In addition to posterior predictive checks, residuals diagnostics can be used to investigate the presence of residual autocorrelation and the viability of the residual normality assumption.
If any of these tests fails, the assumed model is likely to be unfit to describe potential outcomes under control in the post-intervention period. In contrast, if the assumed model is deemed plausible, it doesn't necessarily mean that Assumption~\ref{ass:model} holds for the chosen model. One crucial aspect of diagnostics tests for \cref{ass:model} is that they only entail the data from the pre-intervention period, so model evaluation can be performed completely separately from effect estimation.
In Section~\ref{subsec:model_eval} we show how we used posterior predictive checks to select and validate the model in the context of our study.
In a related context, \cite{antonelli2020heterogeneous} proposed an alternative approach to model evaluation which pertains to using the observed pre-intervention data to simulate hypothetical policies and evaluating the model based on the estimation of their effects.

\subsection{Estimating the causal effect}

Once a model that passes the model assessment step is chosen, causal effect estimation and inference is straightforward by combining the observed outcomes with the samples from the posterior distribution of the imputed potential outcomes $Y^{0t}$.  Let $Y^{0t}_{(b)}$ denote the $b^{th}$ posterior sample for $b = 1, 2, \dots, B$ and $t \geq t^*$. Then, $\big\{ \Delta_{k(b)} = Y_{t^* + k} - Y^{0(t^* + k)}_{(b)} \big\}_{b = 1}^B$ represent $B$ samples from the posterior distribution of the point-wise effect $\Delta_k$ in \cref{eq:effect_at_t}. These estimates are aggregated over the $K$ time points in the post-intervention period to acquire posterior samples of $C\Delta_K$ in \cref{eq:cum_effect}.
Then, we can use the posterior mean as the point-estimator of the causal effects $\Delta_k$ and $C\Delta_K$, and the quantiles of the posterior distribution for inference.

\section{Evaluation of the Hospital Readmissions Reduction Program}
\label{sec:application}

\subsection{Data sources and construction of the data set} 
\label{sec:data}

We collected monthly level death counts from pneumonia, acute myocardial infarction (AMI) and congestive hearth failure (CHF) among people aged more than 65 years old (all gender, all races, all origins) from January 2000 to December 2019. The data were gathered from the Centers for Disease Control and Prevention WONDER database by specifying the underlying cause of death using ICD-10 codes (J12--J18 for pneumonia, I21 for AMI, and I50 for CHF).
We considered death counts at the national level, but we also performed regional-level analyses to study how treatment effects varied across regions. We used the geographical regions corresponding to the Northeast, Midwest, South and West United States as coded by the US Census Bureau (see Supplement \ref{supp_sec:regions} in the Online Resource for the definition of each region).
By analyzing regions separately, we assume that there is no interference across regions, an assumption that is realistic considering the size of each of these regions.

To improve the prediction accuracy of the outcome in the absence of intervention, we also included a set of covariates that might be linked with the outcome, they are unaffected by the intervention, and do not necessarily follow predictable temporal trends. First of all, we considered weather information:
\begin{enumerate}[label=(\arabic*)]
\item Temperature (TEMP) measured as a monthly average in each region, obtained from the National Centers for Environmental Information;
\item Heat index (HEAT) defined as an indicator variable that takes value $1$ when the maximum temperature registered in a given month and region is above $28^{\circ}$C and $0$ otherwise; and
\item Cold index (COLD) defined indicator variable that takes value $1$ when the minimum temperature registered in a given month and region is below $0^{\circ}$C and $0$ otherwise.
\end{enumerate}
The heat and cold indices are included in the model in addition to temperature as extremely hot or cold weather is believed to be a predictor of mortality \citep{medina2006extreme}, and its occurrence, though temporally correlated, does not always exhibit a stable pattern over time. We also included
\begin{enumerate}[label=(\arabic*), resume]
\item An air pollution metric reflecting yearly averages of  PM$_{2.5}$ concentrations at the national level obtained from the US Environmental Protection Agency; and
\item A poverty measure (POV) defined as the percentage of population below the poverty level, measured yearly both at the regional and national level, obtained from the Census Bureau. 
\end{enumerate}
{Since we are evaluating a possible impact of the HRRP on mortality, we excluded information on admitted patients such as comorbidities, since whether a patient is admitted or not, and as a result their characteristics, might be affected by the intervention \citep{casalino2007will}.}

\subsection{Causal assumptions and model evaluation for estimating the effects of the HRRP}
\label{subsec:model_eval}

Even though the financial penalties introduced by the HRRP started to be applied in 2012, we set the intervention date $t^*$ as March 2010, corresponding to the time period that the policy reform was passed.
We use the passage of the law instead of its implementation as the time of intervention to account for the possibility that hospitals might have changed their behaviour right after the program's announcement, and in expectation of the program's implementation (\cref{ass:no_anticipation}). Therefore, our data include a time period of approximately 10 years pre-intervention, and 9 years post-intervention. We return to evaluating \cref{ass:no_anticipation} in the end of this section, and after model choice according to \cref{ass:model} has been completed.

Since the covariates considered are exogenous to the health system, we expect that they are unaltered by the intervention (\cref{ass:model}).
While temperature data may follow cyclical patterns, the poverty measure reverts its increasing trend shortly after the HRRP (see Figure \ref{fig:ts_covariates}). Thus, by including it in our models we allow for changes in our outcome series in the post intervention period due to the trend change in poverty levels, which would otherwise not be possible to predict, as explained in Section \ref{subsec:confounding}.

To predict the counterfactual outcome in the post-intervention period, we estimate the outcome model process based on data from the control, pre-intervention period, and using Bayesian Structural Time Series models \citep{West:Harrison:2006}. 
These models require a preliminary analysis to characterize the features of the generating process (e.g., linear trend, seasonality, cycle). In this specific case, the data exhibit a yearly seasonal pattern, which is visible from the autocorrelation functions in Figure \ref{fig:acf} of the Online Resource. Similar seasonal patterns are observed across the three conditions, and both at a national and regional level (see Figures \ref{fig:ts} and \ref{fig:ts_regional} in the Online Resource). The outcome evolution in the pre-intervention period exhibits decreasing trends for pneumonia and AMI (see Figure \ref{fig:ts} in the Online Resource).
We formally investigated whether a local level or a local linear trend best describes these trends in the outcome process based on the posterior predictive checks described in \cref{subsec:plausibility_assumpt}. We also investigated whether the exogenous covariates and seasonality should be included in the model or not in a similar manner. The results of the model evaluation process for all models considered are included in the Online Resource. Based on these results, we find that, for each of the three outcomes, a local linear trend model with covariates and a seasonal component fits the observed, pre-intervention outcome data the best. This model can be written as
\begin{equation}
\begin{aligned}
Y_t & = \mu_t + \gamma_t + X_t \beta + \varepsilon_t & \varepsilon_t \sim N(0, \sigma^2_{\varepsilon})\\
\mu_{t+1} & = \mu_t + \nu_t + \eta_{t, \mu} & \eta_{t,\mu} \sim N(0, \sigma^2_{\mu})\\
\nu_{t+1} & = \nu_t + \eta_{t, \nu} & \eta_{t,\nu} \sim N(0, \sigma^2_{\nu})\\
\gamma_{t+1} & = - \sum\limits_{s=0}^{S-2} \gamma_{t-s} + \eta_{t,\gamma} & \eta_{t,\gamma} \sim N(0, \sigma^2_{\gamma})
\end{aligned}
\label{eqn:model}
\end{equation}
where $Y_t$ is the outcome time series; $X_t$ is a vector of covariates with coefficients $\beta$; $\mu_t$, $\nu_t$ and $\gamma_t$ are, respectively, the trend component, the random slope and the seasonal component; and $\varepsilon_t, \eta_{t, \mu}, \eta_{t, \nu}, \eta_{t, \gamma}$ are serially independent random variables with zero mean and constant variances. To understand the contribution of each component, consider a simple case with no covariates and no seasonal term: if $\nu_{t} = 0$, the trend component evolves as a random walk and model (\ref{eqn:model}) reduces to a local level; instead, if $\eta_{t,\mu} = \eta_{t,\nu} = 0$, then $\nu_{t+1} = \nu_t = \nu$ and $\mu_{t+1} = \mu_t + \nu$, the trend would be exactly linear making (\ref{eqn:model}) a deterministic linear trend plus noise model. 
Therefore, a local linear trend model allows for flexible, non-linear temporal trends in the outcome time series, that could capture predictors of mortality whose long term trends persist during our study period (see \cref{subsec:confounding}).
Finally, the seasonal component $\gamma_t$ allows each season to have a different contribution to the overall mean, while ensuring that over $S$ seasons the aggregate contribution of $\gamma_t$ is centered at zero. 
This model is also used for the regional analyses.
Even though models with Gaussian errors for ITS analyses do not perform well for count data in general \citep{ye2022comparison}, it is expected that it is a sufficiently good approximation here since the observed counts are very large \citep{Agresti:2013, Coxe:West:Aiken:2009}, and our posterior predictive checks illustrate good performance in predicting the observed counts (first column in \cref{fig:ppc_llinear}).
In \cref{subsec:other_analyses} we consider alternative model specifications and discuss how the results from these alternative analyses relate to the one presented here.

With this model choice, we return to evaluating the presence of anticipation effects in \cref{ass:no_anticipation} following the instructions given in Section \ref{subsec:plausibility_assumpt}. More specifically, we fictionally set the intervention date earlier on December 2009 and then forecast what would have happened in the following three months. We estimate anticipatory effects for all three conditions with 95\% credible intervals overlapping zero (effects represent number of deaths avoided or caused by the HRRP; pneumonia: -1,323 with 95\% CI from -4,112 to 1,329; AMI: -1,344 with 95\% CI from -3,997 to 1,338; and CHF: 579 with 95\% CI from -390 to 1,451). Therefore, there is no indication that \cref{ass:no_anticipation} is violated.



\subsection{Evaluation of the impact of the HRRP on mortality}
\label{subsec:program_evaluation}

\begin{figure}[!t]
\centering
\caption{Monthly death counts from pneumonia, AMI and CHF at the national level. The dashed vertical bar indicates the intervention date.}
\label{fig:ts}
\includegraphics[scale=0.55]{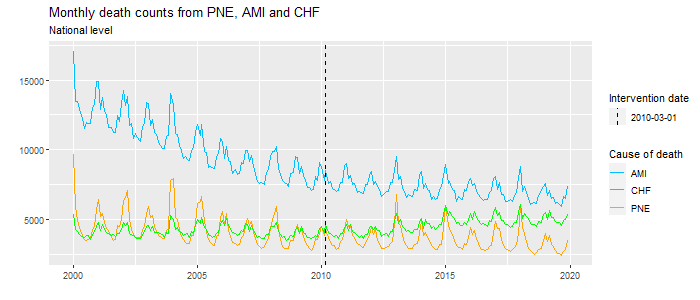}
\end{figure}

We estimated the effect of the passage of the HRRP on mortality from pneumonia, AMI and CHF among the elderly at the national level and in each of the four regions, and at three different time horizons after the intervention: short-term (March 2010--December 2012), mid-term (March 2010--December 2015), and long-term (March 2010--December 2019). For the regional analyses, we estimated the model on mortality data from pneumonia, AMI and CHF for each region separately. We excluded the  PM$_{2.5}$ concentration due to lack of easily-available regional data.\footnote{Looking at Figure \ref{fig:inclusion_probs} in the Online Resource, we believe that  PM$_{2.5}$ can be excluded from the regional analysis without concerns. Indeed, the inclusion probabilities of PM$_{2.5}$ are small in all models; moreover, the sensitivity analysis performed at the national level shows robustness to different choices of predictors (namely, the effects in Table \ref{tab:res}  are in line with the estimates resulting from a local linear trend and seasonal model without PM$_{2.5}$, see panel (1) in Table \ref{tab:sens}).}
All the computations were done using the \texttt{CausalImpact} R package \citep{Brodersen2015}. 
Figure \ref{fig:ts} shows the evolution of the raw death counts from the three conditions at the national level. The presence of seasonal patterns in the considered time series are evident.

\begin{figure}[!t]
\centering
\caption{Cumulative effect of the HRRP on mortality from pneumonia, AMI and CHF at the national and regional level for the intermediate time horizon ending in December 2015. Plots for the remaining regions are provided in the Online Resource.}
\label{fig:res}
\includegraphics[scale=0.45,trim=5 250 0 30, clip]{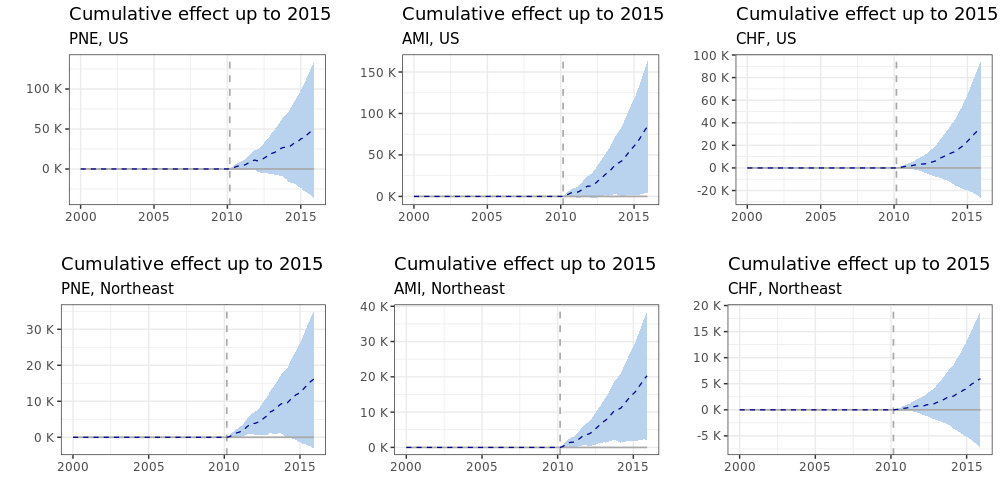} \\
\includegraphics[scale=0.45,trim=0 0 0 280, clip]{cum_forecast_select.png}
\end{figure}

\cref{tab:res} presents the results of our national analysis in terms of estimates of C$\Delta_K$ and corresponding 95\% credible intervals for the short-, mid-, and long-term.
In \cref{tab:res_regional} we provide estimates of C$\Delta_K$ for the four regions and the three time horizons. To ease cross-regional comparisons, we also provide estimates and 95\% credible intervals for the proportion of deaths by each condition that is attributed to the HRRP (by diving our estimates and credible interval bounds with the observed number of deaths in each region).
Lastly, \cref{fig:res} shows the cumulative effect of the HRRP across the US and in the Northeast for each condition and until December 2015 (short to mid-term). The plots for the other regions can be found in the Online Resource.

At the national level, we find that the HRRP led to a medium and long-term increase in mortality from AMI.
By comparing the national results with those emerging from the regional break down, we obtain interesting findings. Short-term mortality is concentrated in the first two regions for pneumonia. Interestingly, even though we did not find the presence of a short run effect for AMI at the national level, we can see mortality from AMI increased due to the HRRP in three regions (the null result in the South US might be behind the absence of a causal effect at the national level).
For the mid-term effects, we can see that the increase in mortality from AMI is mostly driven by the Northeast, Midwest and West regions. Finally, the results from our long-term regional analysis show that the national result for AMI is mostly explained by the Northeast and West US regions. Finally, both the national and regional analyses reflected null effect estimates of the HRRP on CHF mortality.

As we see in \cref{fig:res}, the uncertainty bands for the cumulative effect estimates get wider as we predict further into the future. This is also true for the point-wise effect estimates (see \cref{fig:regional_predicted}). This is due to the auto-regressive mean structure which requires uncertainty quantification for outcomes further in the future to incorporate the uncertainty of parameters and imputed potential outcomes for all intermediate time periods. We find this to be an attractive property of this approach since it represents that the observed outcome time series trends are more trustworthy when extrapolating in the near future. For a fixed $k$, the width of the uncertainty band for our causal effect estimators will not go to zero, irrespective of the length of the pre-intervention period. That is because we estimate the causal effect for a single unit, we do not benefit from estimating averages over a sample, and as a result the uncertainty embedded in the posterior predictive distribution for the unit's missing potential outcome will propagate to our estimator. Even though including predictors of the outcome could reduce the outcome model's residual variance and as a result the variance of the imputed potential outcomes, the potential outcomes' inherent variability will persist.

\begin{table}[!t]
\centering
\caption{Cumulative causal effect and $95 \%$ credibility intervals of the HRRP on mortality from pneumonia, AMI and CHF at the national level. In this table, C$\Delta$ denotes the cumulative impact at the end of three different time horizons (short, mid, and long-term). Results with 95\% CIs that do not include 0 are shown in bold.} 
\label{tab:res}
\begin{adjustbox}{width=1\textwidth}
\begin{tabular}{rrrrrrrrrrrr}
  \hline \hline
 & \multicolumn{3}{c}{PNE} & & \multicolumn{3}{c}{AMI} & & \multicolumn{3}{c}{CHF} \\
 \cline{2-12}
 & C$\Delta$ & $2.5 \%$ & $97.5 \%$ & & C$\Delta$ & $2.5 \%$ & $97.5 \%$ & & C$\Delta$ & $2.5 \%$ & $97.5 \%$ \\ \cline{2-4} \cline{6-8} \cline{10-12}
short-term & 10,795 & -2,092   & 23,665  & & 12,333      & -734      & 25,239      & & 3,780  & -3,035  & 10,509 \\ 
mid-term   & 49,445 & -36,549  & 134,641 & & \bf 83,970  & \bf 3,555 & \bf 162,972 & & 35,300 & -26,811 & 94,469 \\ 
long-term  & 98,849 & -149,887 & 340,540 & & \bf 223,041 & \bf 2,647 & \bf 442,617 & & 94,727 & -96,895 & 27,8605 \\  
\hline \hline
\end{tabular}
\end{adjustbox}
\end{table}

\begin{table}[!t]
\centering
\caption{Causal effect of the HRRP on mortality across US regions. C$\Delta$ denotes the number of deaths attributed to the HRRP. C$\Delta^p$ shows the proportion of deaths attributed to the HRRP (and 95\% credible intervals) in each region. Results are shown for the three time horizons: i) 1 year (December 2011); ii) 5 years (December 2015); iii) 9 years (December 2019). Bold face is used for estimates whose credible interval does not include zero.}
\label{tab:res_regional}
\begin{adjustbox}{width=1\textwidth}
\begin{tabular}{rcrlrlrl}
  \hline \hline
& & \multicolumn{2}{c}{PNE}  &  \multicolumn{2}{c}{AMI}  &  \multicolumn{2}{c}{CHF} \\
 \cline{3-8}
& & C$\Delta$ & C$\Delta^p$ (95\% CI) &  C$\Delta$ & C$\Delta^p$ \ (95\% CI)  &  C$\Delta$ & C$\Delta^p$ \ (95\% CI) \\ 
  \hline
  \parbox[t]{2mm}{\multirow{4}{*}{\rotatebox[origin=c]{90}{short-term}}} 
& Northeast & \bf 3,931 & \bf 22.2 (5.6, 38.5) & \bf 3,541 & \bf 10.7 (1.4, 20.1) & 741 & 4.4 (-4.7, 13.2) \\ 
& Midwest   & \bf 3,517 & \bf 19.4 (1.4, 37.5) & \bf 3,644 & \bf \phantom{0}8.9 (0.2, 17.5) & 995 & 4.2 (-4.4, 12.7) \\ 
& South     & 3,669 & 13.5 (-2.4, 29.1) & 3,394 & \phantom{0}5.3 (-2.3, 12.7) & 582 & 1.7 (-5.8, 9.3) \\ 
& West      & 2,131 & 13.4 (-3.2, 30.1) & 2,270 & \bf \phantom{0}7.8 (0.9, 14.5) & 1,319 & 8.7 (-0.1, 17.5) \\ 
   \hline
  \parbox[t]{2mm}{\multirow{4}{*}{\rotatebox[origin=c]{90}{mid-term}}} 
& Northeast & 16,155 & 28.9 (-5.4, 62.6) & \bf 20,294 & \bf 20.3 (2, 38.7) & 5,958 & 10.3 (-12.8, 32.8) \\ 
& Midwest   & 18,957 & 32.6 (-3.8, 69.4) & \bf 23,100 & \bf 18.4 (1.2, 35.4) & 10,157 & 12.5 (-10.6, 33.6) \\ 
& South     & 16,954 & 19.3 (-13.9, 51.8) & 28,313 & 14.1 (-0.7, 28.9) & 11,898 & 10.1 (-9.1, 28.9)  \\ 
& West      & 14,247 & 27.7 (-8.2, 63.4) & \bf 18,404 & \bf 20.2 (6.1, 34.7) & 9,795 & 19\phantom{.5} (-5.4, 43.1) \\ 
\hline
  \parbox[t]{2mm}{\multirow{4}{*}{\rotatebox[origin=c]{90}{long-term}}} 
& Northeast & 31,269 & 34.2 (-28.5, 95.2) & \bf 52,046 & \bf 33\phantom{.5} (0.2, 65.6) & 14,865 & 14.7 (-26.9, 55.6) \\ 
& Midwest   & 41,004 & 44.2 (-20.9, 110.7) & 57,955 & 28.5 (-1.3, 58) & 25,984 & 18.1 (-21.6, 56.6) \\
& South     & 35,671 & 24.8 (-33.8, 82.5) & 77,855 & 23.4 (-2.6, 49.6) & 32,349 & 15.5 (-18.1, 48.9) \\ 
& West      & 33,938 & 40.2 (-22.7, 102.8) & \bf 55,022 & \bf 36.1 (10.8, 62.1) & 27,003 & 28.6 (-14, 71.2) \\ 
\hline \hline
\end{tabular}
\end{adjustbox}
\end{table}

\subsection{Additional analyses of our data}
\label{subsec:other_analyses}

Considering the policy importance of our results, we performed several checks to evaluate the performance of our approach and the robustness of our qualitative results to the model chosen and the methodology used. These checks correspond to analyses of the effect of the HRRP under the following general categories:
\begin{enumerate*}[label=(\arabic*)]
\item analyses based on our model and alternative choice of predictors,
\item analyses on control time series that are expected to be unaffected by the intervention,
\item analyses based on alternative methodologies that employ a different set of causal assumptions, and
\item analyses based on transformations of our data,
\end{enumerate*}
The analyses are discussed in detail in \cref{supp_sec:robustness_checks}, and the acquired results are summarized here.

First, we evaluate the robustness of our quantitative results to alternative choices of predictor variables. We follow a three-step procedure: We establish alternative sets of predictor variables that one could consider. Then, for each set of predictors and without looking at the effect estimates based on each model, we inspect the posterior predictive checks of the model fits for each outcome and each set of covariates. If the check shows that the model is not a good fit for the observed data, the corresponding set of covariates is discarded. Using the models with predictors that passed the posterior predictive checks, we estimate the causal effects, and we compare them to the results from our original analysis.
For our study, we considered three alternative sets of predictors:
\begin{enumerate*}[label=\alph*)]
\item no predictors;
\item all predictors excluding PM$_{2.5}$; and
\item only weather covariates.
\end{enumerate*}
The posterior predictive checks of the model without predictors (Figure \ref{fig:ppc_no_regress} in the Online Resource) show high residual autocorrelation, and therefore this model is discarded. Conversely, the model without PM$_{2.5}$ and the model with only the weather covariates fit the data well (see Figures \ref{fig:ppc_no_pm25}, \ref{fig:ppc_only_weather} and \ref{fig:trace_no_pm25}, \ref{fig:trace_only_weather} in the Online Resource). The causal effect estimates from these two models are shown in \cref{tab:sens} and they are extremely similar to our original results in \cref{tab:res}, illustrating that our results are robust to the choice of predictor variables.

\begin{table}[!b]
\centering
\caption{Cumulative causal effect and $95 \%$ credibility intervals of the HRRP on mortality from pneumonia, AMI and CHF at the national level from two local linear trend models with different choices of regressors: (1) all regressors excluding PM$_{2.5}$; (2) only weather covariates.}
\label{tab:sens}
\begin{adjustbox}{width=1\textwidth}
\begin{tabular}{lrrrrrrrrrrrr}
  \hline \hline
& & \multicolumn{3}{c}{PNE} & & \multicolumn{3}{c}{AMI} & & \multicolumn{3}{c}{CHF} \\
 \cline{3-13}
& & C$\Delta$ & $2.5 \%$ & $97.5 \%$ & & C$\Delta$ & $2.5 \%$ & $97.5 \%$ & & C$\Delta$ & $2.5 \%$ & $97.5 \%$ \\ \cline{3-5} \cline{7-9} \cline{11-13}
\multirow{3}{*}{(1)} 
& short-term  & 10,756     & -1,814     & 23,048  & & 12,386     & -372      & 25,274  & & 3,432      & -3,244     & 9,906   \\
& mid-term    & 47,828     & -36,526    & 133,777 & & \bf 86,054     & \bf 6,236     & \bf 163,997 & & 31,374     & -32,046    & 91,649  \\
& long-term   & 98,180     & -146,105   & 342,845 & & 216,877    & -5,393     & 437,220 & & 91,494     & -96,189    & 273,341 \\ \hline
\multirow{3}{*}{(2)} 
& short-term  & 10,555     & -1,511     & 22,753  & & 12,570     & -583      & 25,396  & & 3,709      & -2,841     & 10,357  \\
& mid-term    & 48,263     & -35,138    & 130,502 & & \bf 84,239     & \bf 3,634     & \bf 162,949 & & 39,010     & -20,062    & 95,792  \\
& long-term   & 105,614    & -127,016   & 342,462 & & \bf 221,273    & \bf 221       & \bf 439,858 & & 98,079     & -84,724    & 274,474 \\
   \hline \hline
\end{tabular}
\end{adjustbox}
\end{table}

As an additional robustness check, we performed falsification tests by repeating the analysis on death conditions that, in principle, should not be affected by the HRRP: if the estimated impact on such conditions is close to zero, we can be more confident on the results obtained for pneumonia, AMI, and CHF. To do so, we acquired monthly level death counts from the Centers for Disease Control and Prevention WONDER database by specifying as underlying cause of death the following codes: A00--B99 (infectious and parasitic diseases); C00--D49 (tumors); D50--D89 (diseases of the blood); F01--F99 (mental disorders); V00--Y99 (external causes of morbidity). Figure \ref{fig:fals_outcomes} in the Online Resource shows the evolution in mortality from each disease from January 2000 to December 2019. We can observe an increasing trend in mortality from all causes in the considered period. The results from each estimated model are reported in Table \ref{tab:res_falsification} and show no evidence of a causal effect for any of the control outcomes.
These results show that if unpredictable changes in mortality trends exist, they are not present in control conditions so we can be more reassured that such changes are also not present in the target conditions. In addition, falsification tests are also important to ascertain the impact of possible co-occurring policies: if no evidence of an effect is found on control conditions, we can be more confident that co-occurring policies had no impact on target conditions as well. Finally, we verified the absence of anticipatory effects on both the true outcomes and on control conditions (see Table \ref{tab:anticipation} in the Online Resources).

\begin{table}[!b]
\centering
\caption{Falsification tests. Cumulative causal effect of the HRRP on control outcomes at three different time horizons.}
\label{tab:res_falsification}
\begin{adjustbox}{width=1\textwidth}
\begin{tabular}{lrrrrrrrrrrr}
  \hline
  & \multicolumn{3}{c}{Short term} & & \multicolumn{3}{c}{Mid term} & & \multicolumn{3}{c}{Long term} \\ \cline{2-4} \cline{6-8} \cline{10-12}
 & C$\Delta$ & 2.5\% & 97.5\% & & C$\Delta$ & 2.5\% & 97.5\% & & C$\Delta$ & 2.5\% & 97.5\% \\ 
  \hline
Infections       & 3,147      & -2,808     & 9,131  & & 20,725     & -33,304    & 76,348  & & 44,187     & -126,529   & 217,690 \\
Tumors           & 6,548      & -3,068     & 16,106 & & 57,050     & -16,135    & 127,480 & & 160,398    & -40,405    & 361,319 \\
Blood diseases   & 1,371      & -1,868     & 4,565  & & 9,457      & -33,576    & 53,331  & & 18,490     & -140,143   & 172,038 \\
Mental disorders & 12,072     & -854       & 24,870 & & 5,739      & -78,555    & 87,538  & & -221,402   & -506,026   & 24,146  \\
External causes  & 5,673      & -578       & 11,833 & & 41,207     & -27,218    & 109,316 & & 117,288    & -108,225   & 344,842 \\
   \hline
\end{tabular}
\end{adjustbox}
\end{table}

As we discussed in \cref{subsec:confounding}, our approach can suffer if there exist structural changes in the temporal trends of the outcome time series after the intervention that are not due to the intervention. To evaluate the robustness of our qualitative analyses, we considered alternative methodologies that aim to address these issues, namely: interrupted time series, difference-in-differences (DiD) and synthetic control methods. 
For the synthetic control approach, we employed the augmented synthetic control approach of \citet{Benmichael:Feller:Rothstein:2021} using mortality from conditions not targeted by the HRRP as control time series, in line with \cite{gaughan2019paying}.
We include a detailed discussion of these methods and corresponding results in Section \ref{supp_sec:robustness_checks} of the Online Resource. All these approaches rely on underlying assumptions that should be carefully addressed in a complete causal analysis.
We find that results based on the ITS approaches in \citet{Miratrix2019simulating}, as well results based on synthetic control methods yield results which are in line in direction and magnitude with the ones shown in Table \ref{tab:res}.

In our main analyses, 
we used Gaussian models for our mortality counts because our interest is in estimating the {\it number} of deaths that were avoided by or attributed to the HRRP. 
We have found that Poisson regression models with autoregressive mean structure do not perform well in predicting future outcomes due to high uncertainty \citep{Wheeler:2018}. We found that our Gaussian model fit the observed data well even though these are count data. As a companion analysis, we considered alternative model specifications for both the true and the control outcomes, namely, a Gaussian Bayesian time series model estimated on the logarithmic transformation of the monthly death counts and a Gaussian Bayesian time series model estimated on the logit transformation of the monthly death rate from each condition. The results are reported in the Online Resource and seem to exclude the presence of causal effects. However, in \cref{supp_sec:robustness_checks} we show that such models target different causal effect estimands: a multiplicative effect on the outcome for the log-normal model, a causal effect on the log-odds ratio for the logit transformation. Thus, such results seem to suggest that the HRRP policy incentive scheme had an additive effect on the death counts from pneumonia and AMI, as shown in Table \ref{tab:res}.

\section{Discussion}
\label{sec:discussion}

We find that the main methodological contributions of this paper are three-fold. In a time-series setting with one time series and one interventional time point, we (1) defined quantities to be estimated in terms of potential outcomes, (2) formalized the assumptions on which a previously-developed~\citep{Brodersen2015} Bayesian time series model was based (which had not been done until now), and importantly (3) provided a comprehensive interrogation of the assumptions underlying the estimation of such causal estimands in the context of our study, laying the foundation for the use of this method in a wide range of applications assessing clinical outcomes for health policy in non-randomized settings.

In our illustrative study, we evaluated the effect of the HRRP passage on mortality among the elderly. One of the key difference of our approach compared to previous associational studies is our focus on the whole region of interest as one unit measured over time, which allows us to estimate the effect of the HRRP passage on the whole population it meant to serve.
Since the presented approach bypasses some complicating aspects of policy evaluation, we view our approach as complementary to approaches that compare penalized to non-penalized hospitals.

To our knowledge, this is the first paper evaluating the causal effect of the HRRP. Previously published analyses have suggested that after enactment of the HRRP, readmissions have decreased for all of the 3 initial penalty conditions~\citep{Zuckerman2016,Desai2016,Wasfy2017}, even though these reductions in readmissions are likely due to increases in post-discharge emergency department visits or observation stays~\citep{Wadhera2019hospital}. 
Although our results are not directly comparable to the above studies, a fraction of reduced readmissions may result in increased mortality; in addition, our analysis is also able to capture the portion of mortality due to the HRRP arising from other sources (e.g., delayed first admissions and in-hospital mortality), or due to the possibility that hospital resources were redistributed away from reducing mortality to reducing readmissions.

Even though reductions in readmissions are encouraging, our analysis raises concerns about increased mortality caused by the HRRP. 
From the perspective of policy evaluation, our results illustrate that the future of the HRRP might have to be reassessed and future amendments might be necessary to avoid a detrimental impact of the policy on the population it was meant to serve.

\section*{Declarations}

\subsection*{Funding}
Funding was provided by National Institutes of Health R01 GM111339, R01 ES024332, R01 ES026217, P50MD010428, DP2MD012722, R01 ES028033, USEPA 83615601, and Health Effects Institute 4953-RFA14-3/16-4. The contents of this work are solely the responsibility of the grantee and do not necessarily represent the official views of the USEPA. Further, USEPA does not endorse the purchase of any commercial products or services mentioned in the publication. 
Research described in this article was conducted under contract to the Health Effects Institute (HEI), an organization jointly funded by the United States Environmental Protection Agency (EPA) (Assistance Award No.CR-83467701) and certain motor vehicle and engine manufacturers. The contents of this article do not necessarily reflect the views of HEI, or its sponsors, nor do they necessarily reflect the views and policies of the EPA or motor vehicle and engine manufacturers.
Dr. Wasfy is supported by National Institutes of Health and Harvard Catalyst KL2 TR001100 and American Heart Association 18CDA34110215.

Dr. Menchetti and Prof. Mealli thank the Department of Excellence 2018-2022 funding provided by the Italian Ministry of Education, University and Research (MIUR).

The authors have no conflicts of interest to declare that are relevant to the content of this article.



\bibliographystyle{plainnat}
\bibliography{Hospitals}


\setcounter{page}{1}
\setcounter{page}{1}
\setcounter{section}{0}    
\renewcommand{\thesection}{\Alph{section}}
\setcounter{equation}{0}    
\renewcommand{\theequation}{S.\arabic{equation}}
\setcounter{table}{0}    
\renewcommand{\thetable}{S.\arabic{table}}
\setcounter{figure}{0}    
\renewcommand{\thefigure}{S.\arabic{figure}}

\newgeometry{right = 1cm, left = 1cm, top = 2cm, bottom = 2cm}

\begin{center}
\Large
Supporting information for the manuscript \\ \thetitle \\ by \\ \theauthor
\end{center}


\section{Census regions and divisions of the United States}
\label{supp_sec:regions}

The regional data gathered from the WONDER database follow the region definition of the Census Bureau. We report below the map and the list of states belonging to each region, as downloaded from the website \url{https://www2.census.gov/geo/pdfs/maps-data/maps/reference/us_regdiv.pdf} (Retrieved November 19, 2021).

\begin{figure}[h!]
    \centering
    \includegraphics[scale = 0.5]{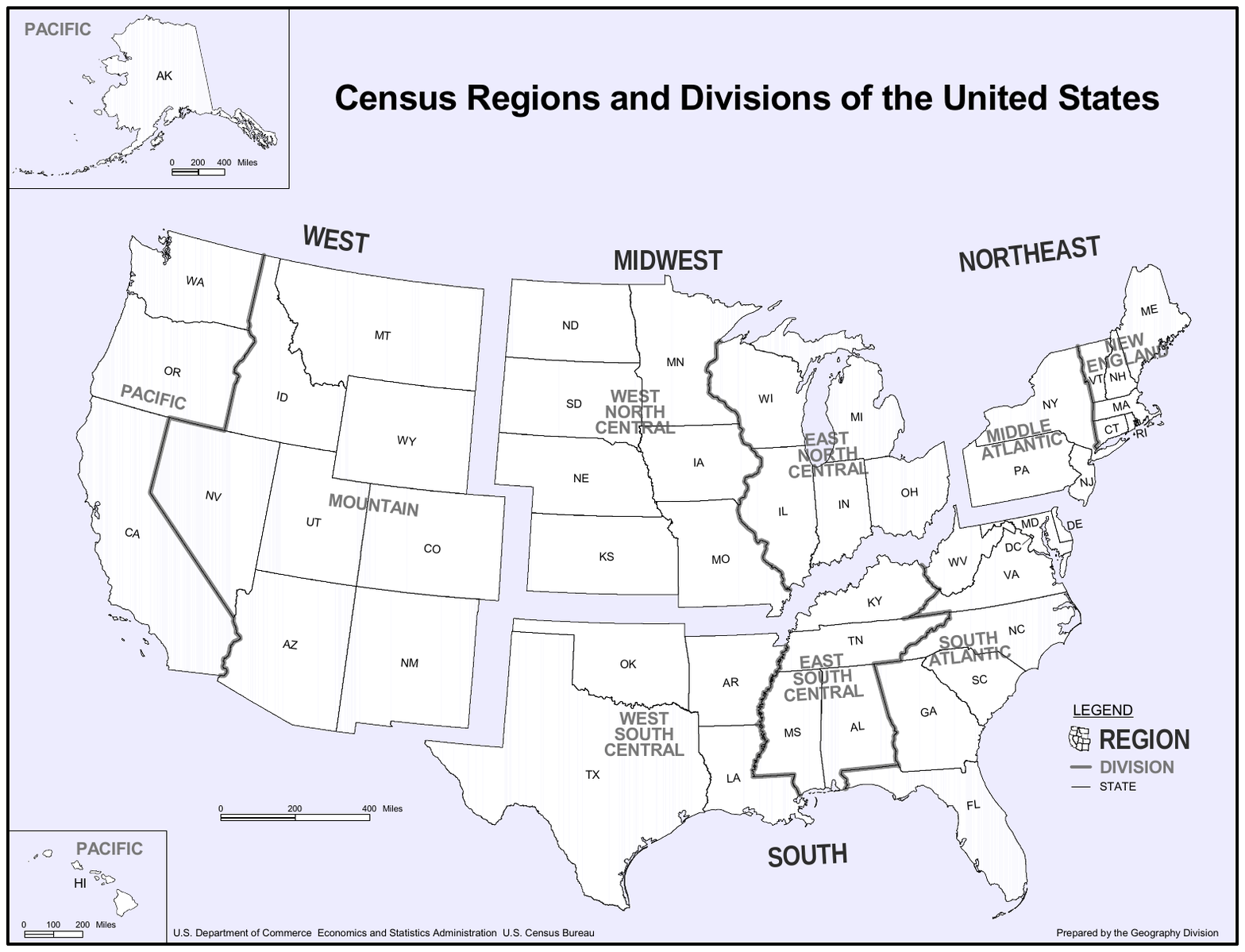}
\end{figure}

\newgeometry{top = 2cm, bottom = 2cm, left = 1.5cm, right = 1.5cm}
\begin{figure}
    \centering
    \includegraphics[width=0.9\textwidth]{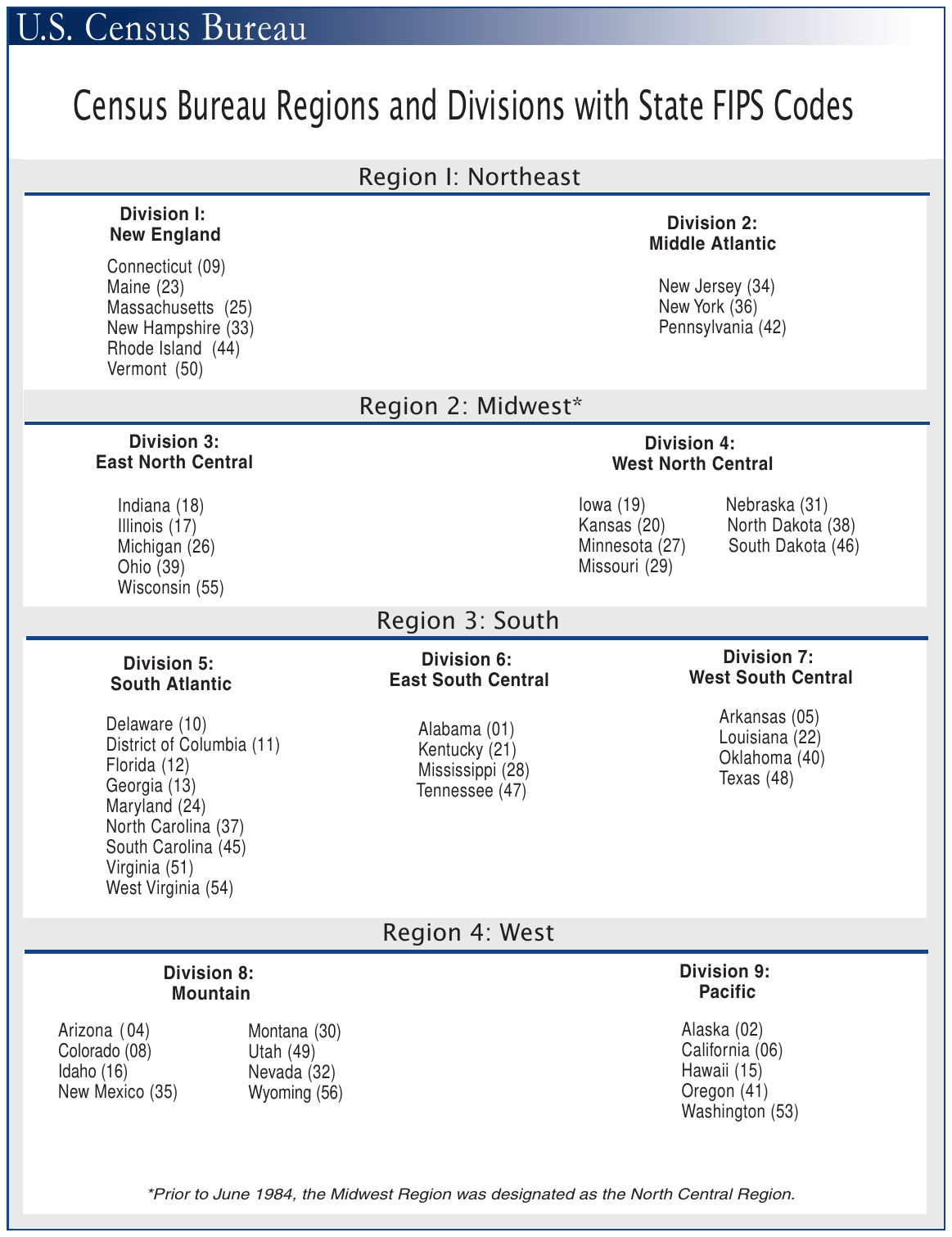}
\end{figure}
\restoregeometry

\clearpage
\section{Investigating patterns in our data}

\begin{figure}[H]
\vspace{20pt}
\centering
\caption{Autocorrelation functions of the three outcome variables (death counts from pneumonia, AMI and CHF) at the national level. All plots show spikes every $12$ lags, indicating a clear seasonal pattern.}
\label{fig:acf}
\includegraphics[scale=0.47]{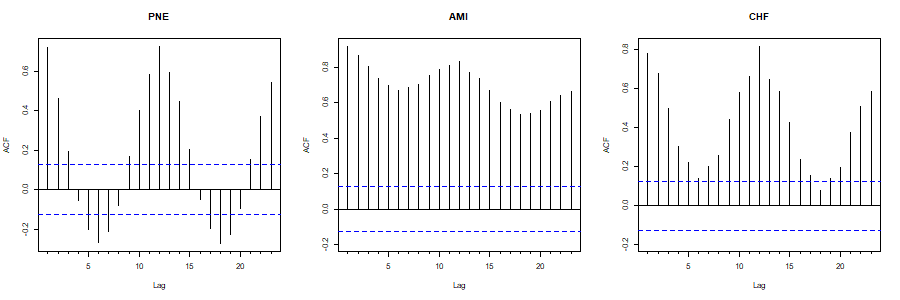}
\end{figure}

\begin{figure}[p]
\centering
\caption{Evolution of the covariates during the analysis period. Starting from top: i) percentage of people below the poverty level; ii) percentage of people below the poverty level by region; ii) concentration of fine particulate matter (yearly average, $\mu g /m^3$); air temperature (monthly average, $^\circ C$) by region.} 
\label{fig:ts_covariates}
\includegraphics[scale=0.46]{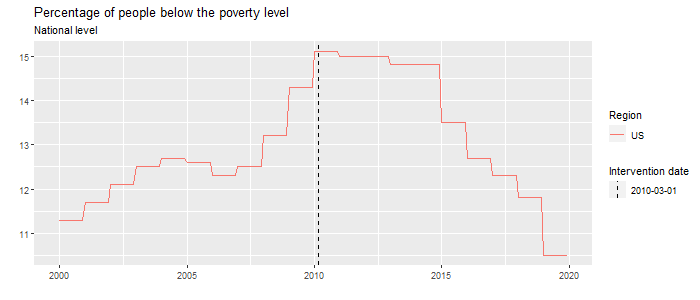} \\
\includegraphics[scale=0.46]{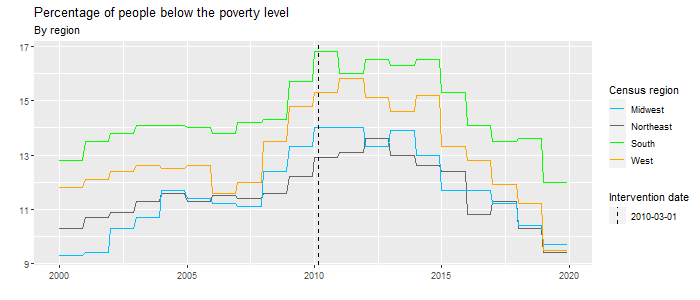} \\
\includegraphics[scale=0.46]{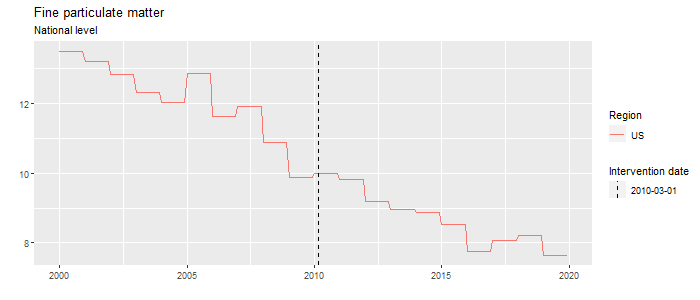} \\
\includegraphics[scale=0.46]{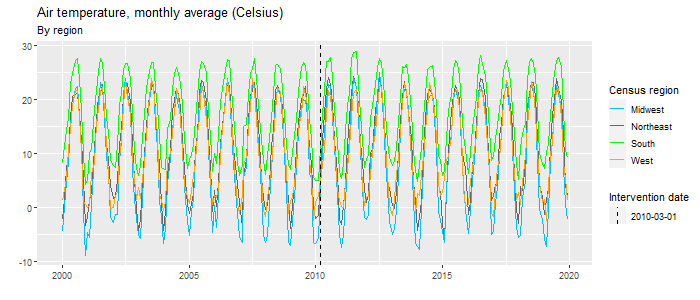}
\end{figure}

We first investigated seasonality in our data, which is evident from \cref{fig:acf} showing the autocorrelation plots for each outcome at the national level: the spikes every 12 lags indicate yearly seasonality. The seasonality patterns are also evident from \cref{fig:ts} in the main paper and from \cref{fig:ts_regional} below, where we also notice a decreasing trend in mortality from pneumonia and AMI both at the national and regional level. Finally, Figure \ref{fig:ts_covariates} shows the evolution of covariates during the analysis period. There, we see that the poverty levels show a change in their trend in the pre- and in the post-intervention period.

\begin{figure}[p]
\centering
\caption{Monthly death counts from pneumonia (panel A), AMI (panel B) and CHF (panel C), by census region. The dashed vertical bar indicates the intervention date and the four census regions are coded as follows: Northeast, Midwest, South, West.}
\label{fig:ts_regional}
\begin{tabular}{m{1cm}m{12cm}}
A) & \includegraphics[scale=0.5]{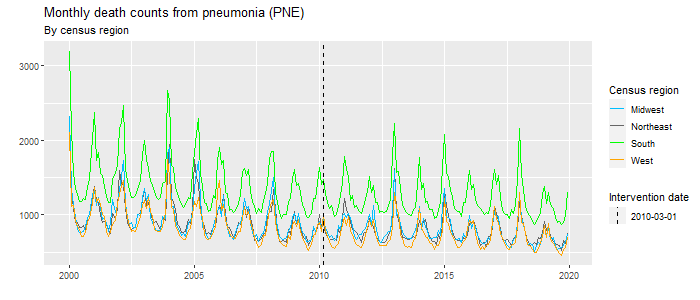} \\
B) & \includegraphics[scale=0.5]{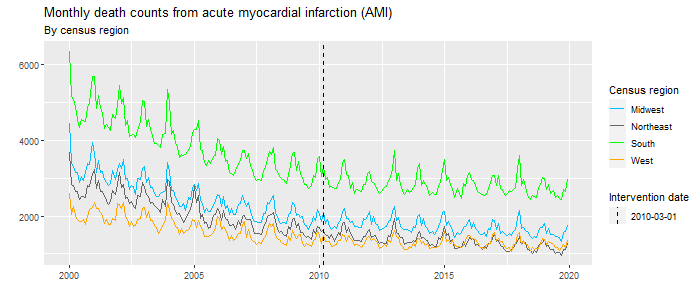} \\
C) & \includegraphics[scale=0.5]{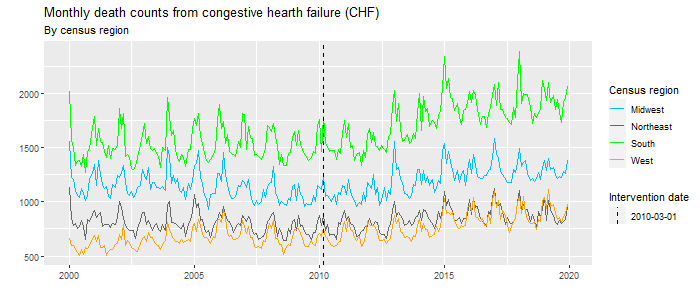} \\
\end{tabular}
\end{figure}

\clearpage

\section{Investigating model fit in the pre-intervention period}
\label{supp_sec:model_fit}

In Section \ref{subsec:plausibility_assumpt} we assert that even though Assumption \ref{ass:model} can not be formally tested, we can still check whether the model fits well to the pre-intervention data: if it doesn't, it is unlikely that the assumed model adequately describes post-treatment outcomes in the absence of the intervention. To this aim, we employ posterior predictive checks, we assess model convergence and, since we include regressors, we check posterior inclusion probabilities to see whether they help predicting the outcome in the pre-intervention period.

\subsection{Posterior predictive checks}
\label{subsec:ppc}

\begin{figure}[p]
\centering
\caption{Posterior predictive checks for the selected \textbf{local linear trend and seasonal models} estimated in the pre-intervention period on the death counts from each condition (PNE, AMI and CHF) at the national level.}
\label{fig:ppc_llinear}
\includegraphics[scale=0.33]{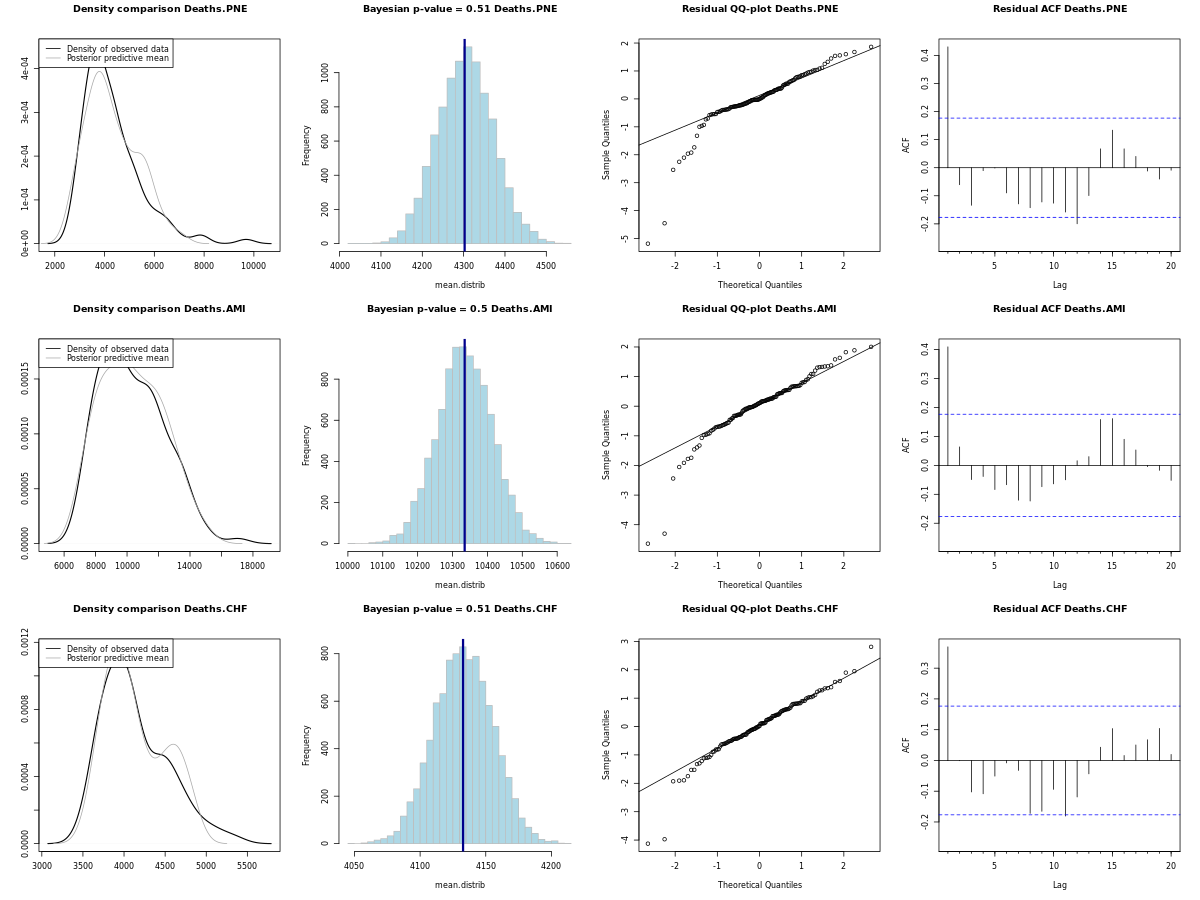}

\vspace{20pt}

\centering
\caption{Posterior predictive checks for a \textbf{local linear trend and seasonal model with no regressors} estimated in the pre-intervention period on the death counts from each condition.}
\label{fig:ppc_no_regress}
\includegraphics[scale=0.33]{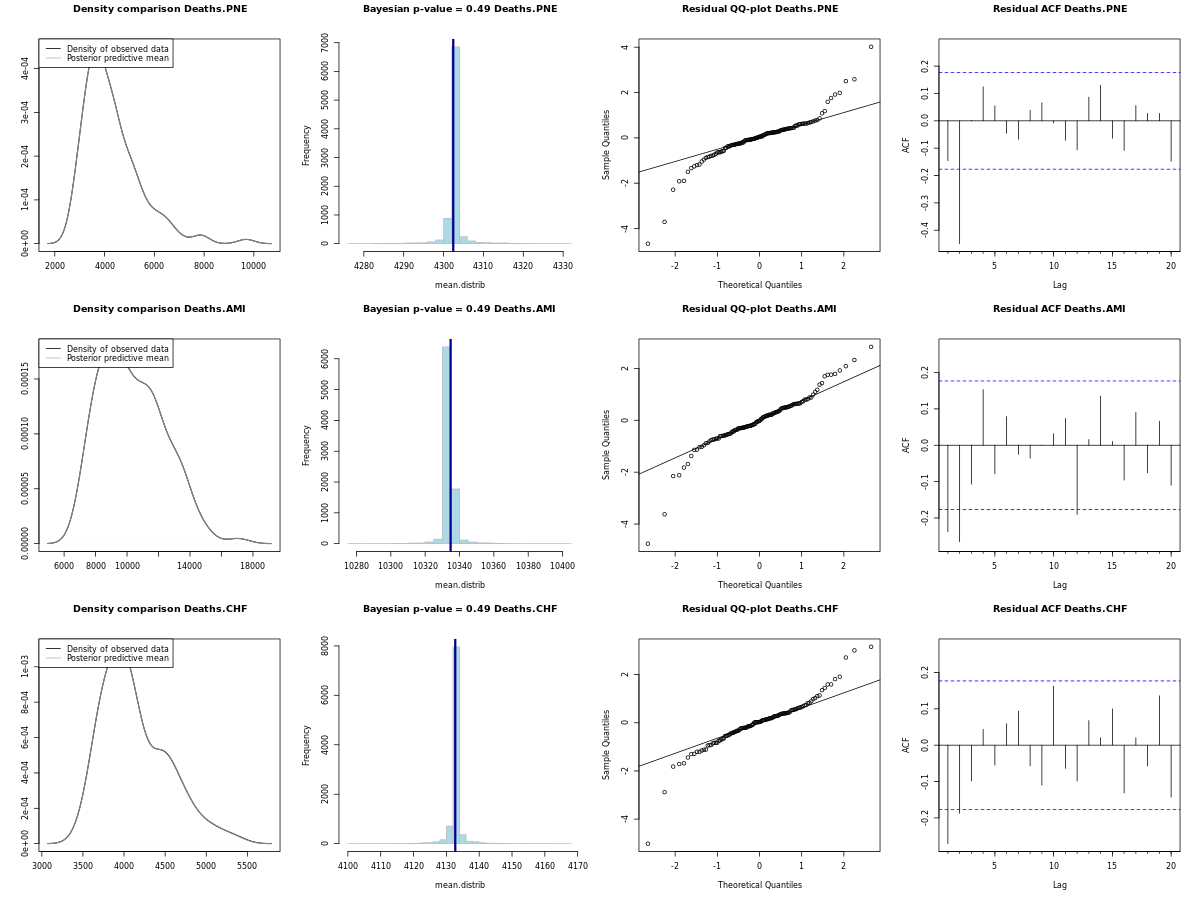}
\end{figure}

\begin{figure}[p]
\centering
\caption{Posterior predictive checks for a \textbf{local level and seasonal model with no regressors} estimated in the pre-intervention period on the death counts from each condition.}
\label{fig:ppc_llevel_nocov}
\includegraphics[scale=0.33]{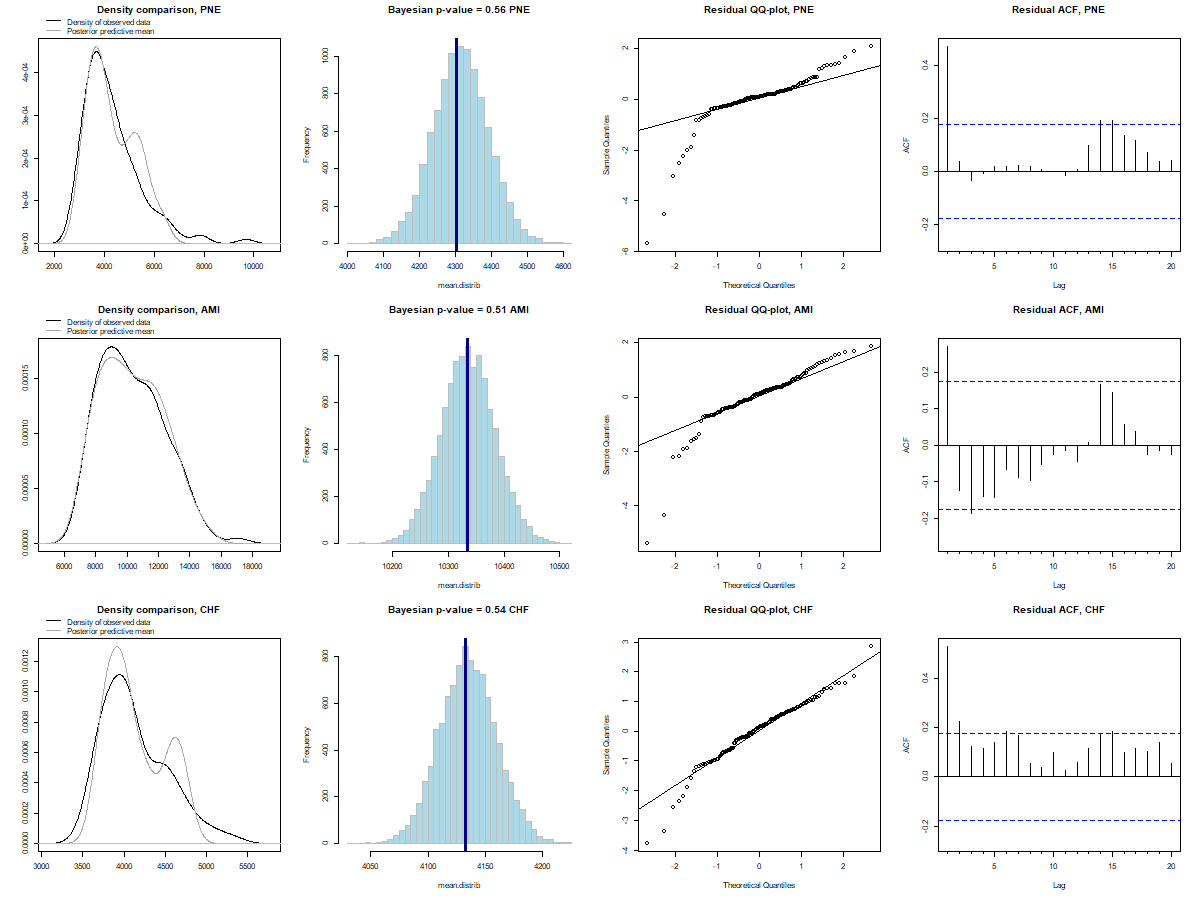}

\vspace{20pt}

\centering
\caption{Posterior predictive checks for a \textbf{local level and seasonal model with regressors} estimated in the pre-intervention period on the death counts from each condition.}
\label{fig:ppc_llevel}
\includegraphics[scale=0.33]{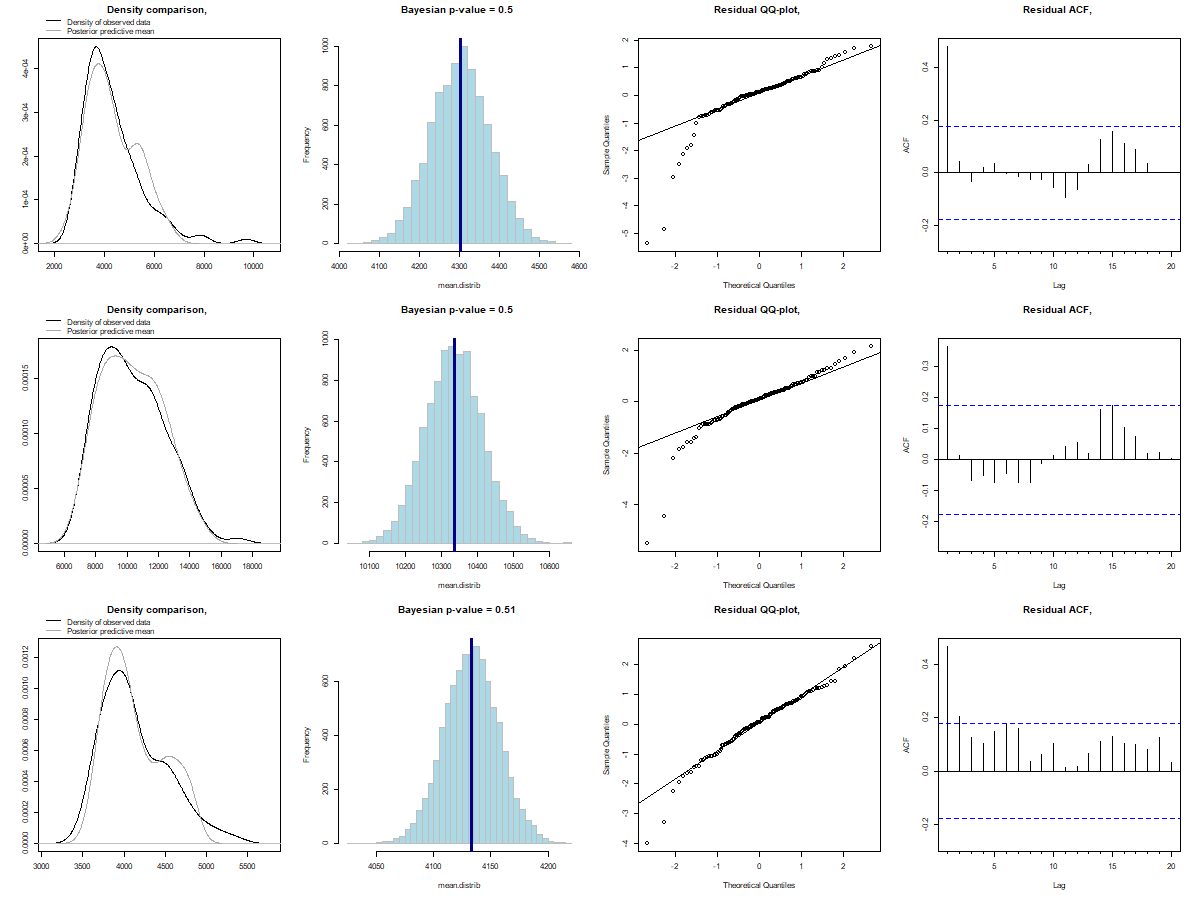}
\end{figure}

\begin{figure}[p]
\centering
\caption{Posterior predictive checks for a \textbf{local linear trend and seasonal model excluding PM$_{2.5}$} from the set of regressors.}
\label{fig:ppc_no_pm25}
\includegraphics[scale=0.33]{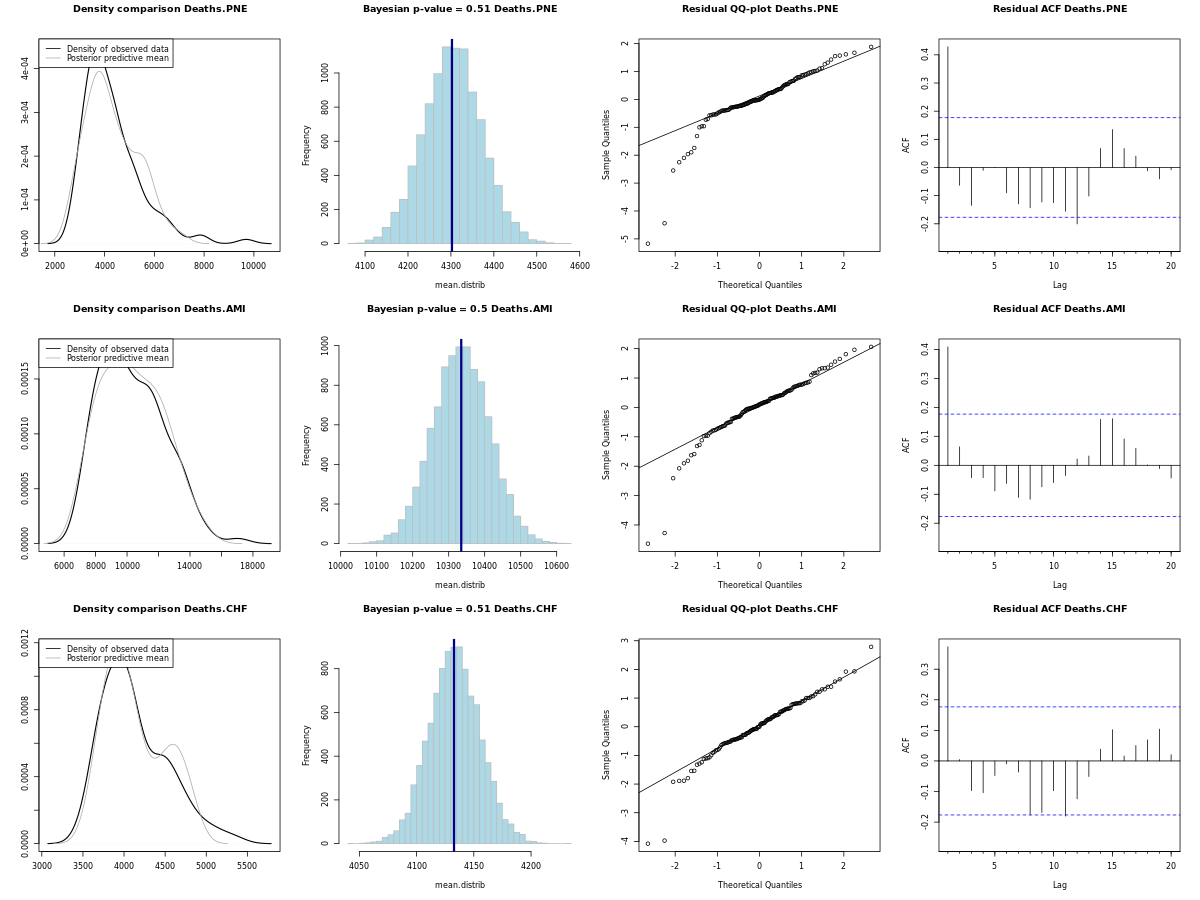} 

\vspace{20pt}

\centering
\caption{Posterior predictive checks for a \textbf{local linear trend and seasonal model having as regressors the weather covariates only}.}
\label{fig:ppc_only_weather}
\includegraphics[scale=0.33]{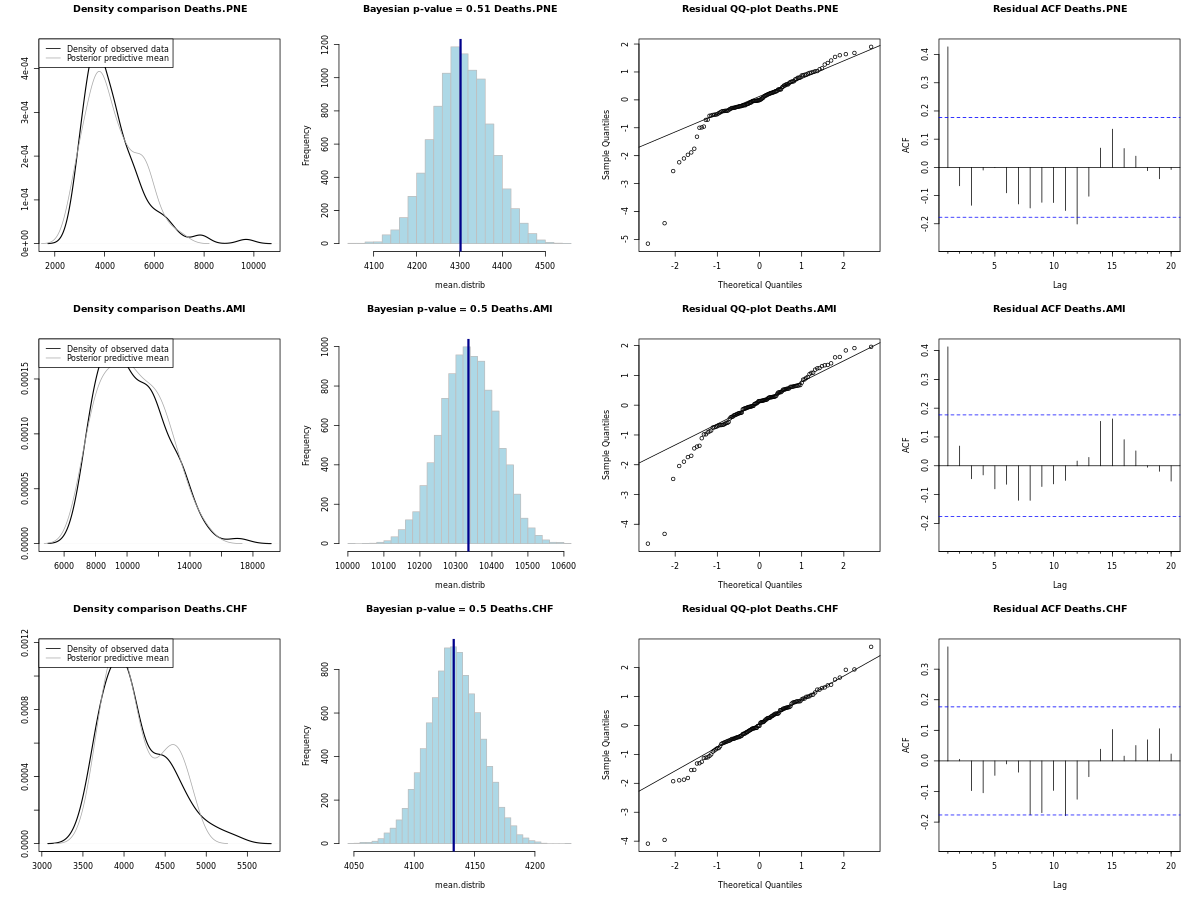} 
\end{figure}

As described earlier, we use posterior predictive checks to inspect the model fit to pre-intervention data. The results from the posterior predictive checks for each model are shown as:

\begin{itemize}
\item \cref{fig:ppc_llinear}: selected model with local linear trend, seasonality, and all predictors.

\item \cref{fig:ppc_no_regress}: model with local linear trend, seasonality,, and no predictors.

\item \cref{fig:ppc_llevel_nocov}: model with local trend, seasonality, and no predictors.

\item \cref{fig:ppc_llevel}: model with local trend, seasonality, and all predictors.

\item \cref{fig:ppc_no_pm25}: model with local linear trend, seasonality, and all predictors excluding PM$_{2.5}$.

\item \cref{fig:ppc_only_weather}: model with local linear trend, seasonality, and only weather-related predictors.

\end{itemize}

We notice that the models without regressors exhibit excess residuals autocorrelation (Figures \ref{fig:ppc_no_regress} and \ref{fig:ppc_llevel_nocov}). The local level model with the addition of regressors shows better fit to pre-intervention data than the model without (Figure \ref{fig:ppc_llevel}) but it still has slightly larger residuals autocorrelations compared to the selected model. 
The posterior predictive checks that better align to those of the selected model are shown in Figures \ref{fig:ppc_no_pm25} and \ref{fig:ppc_only_weather}, corresponding to the local linear trend and seasonal models with only a subset of regressors. 

\clearpage
\newgeometry{top = 1.5cm, bottom = 1.5cm, left = 1.5cm, right = 1.5cm}

\restoregeometry
\restoregeometry


\clearpage

\subsection{MCMC convergence}
\label{subsec:convergence}

\begin{figure}[h!]
\centering
\caption{Trace plots for the selected \textbf{local linear and seasonal model} estimated in the pre-intervention period on the death counts from each condition (PNE, AMI and CHF) at the national level.}
\includegraphics[scale=0.38]{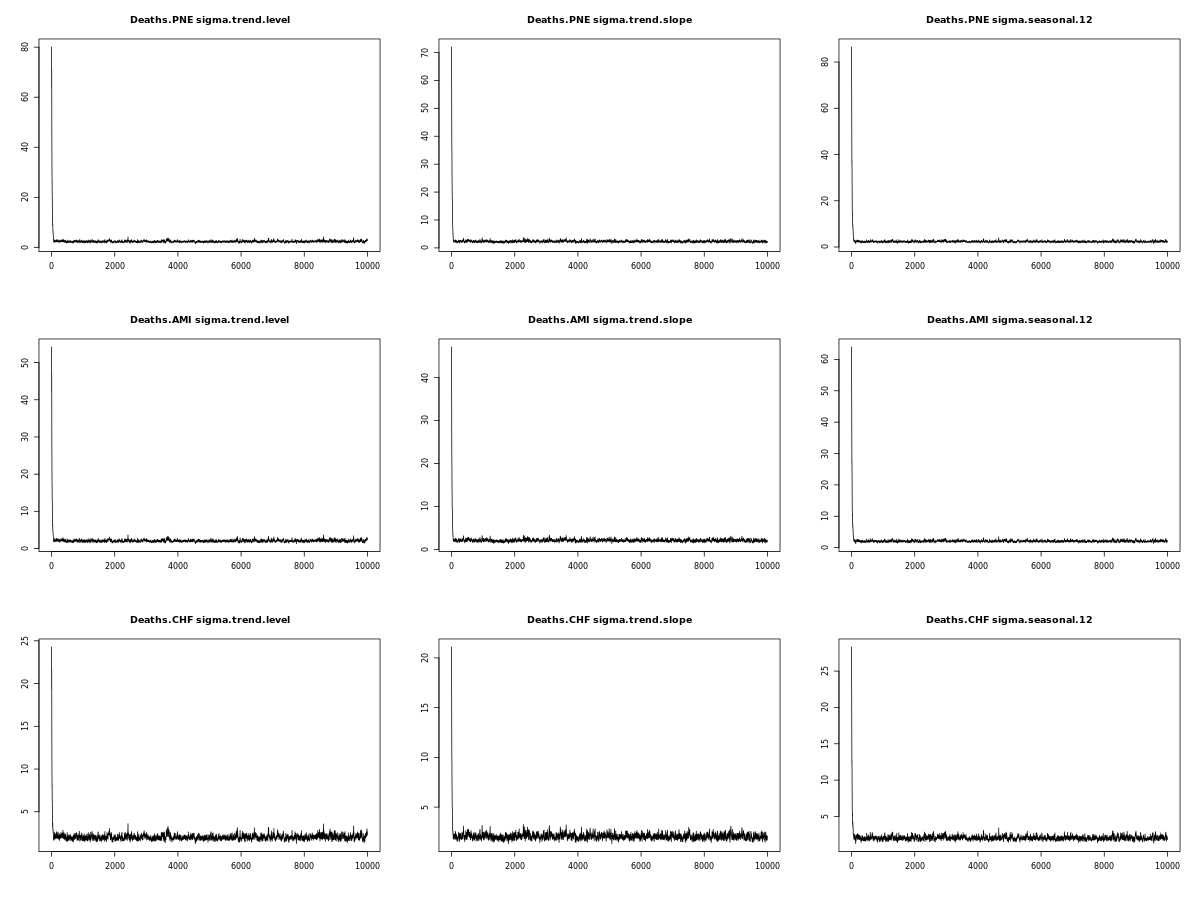}
\label{fig:trace_llinear}
\end{figure}

\begin{figure}[p]
\centering
\caption{Trace plots for the alternative \textbf{local linear and seasonal model without regressors} estimated in the pre-intervention period on the death counts from each condition.}
\label{fig:trace_llinear_nocov}
\includegraphics[scale=0.3]{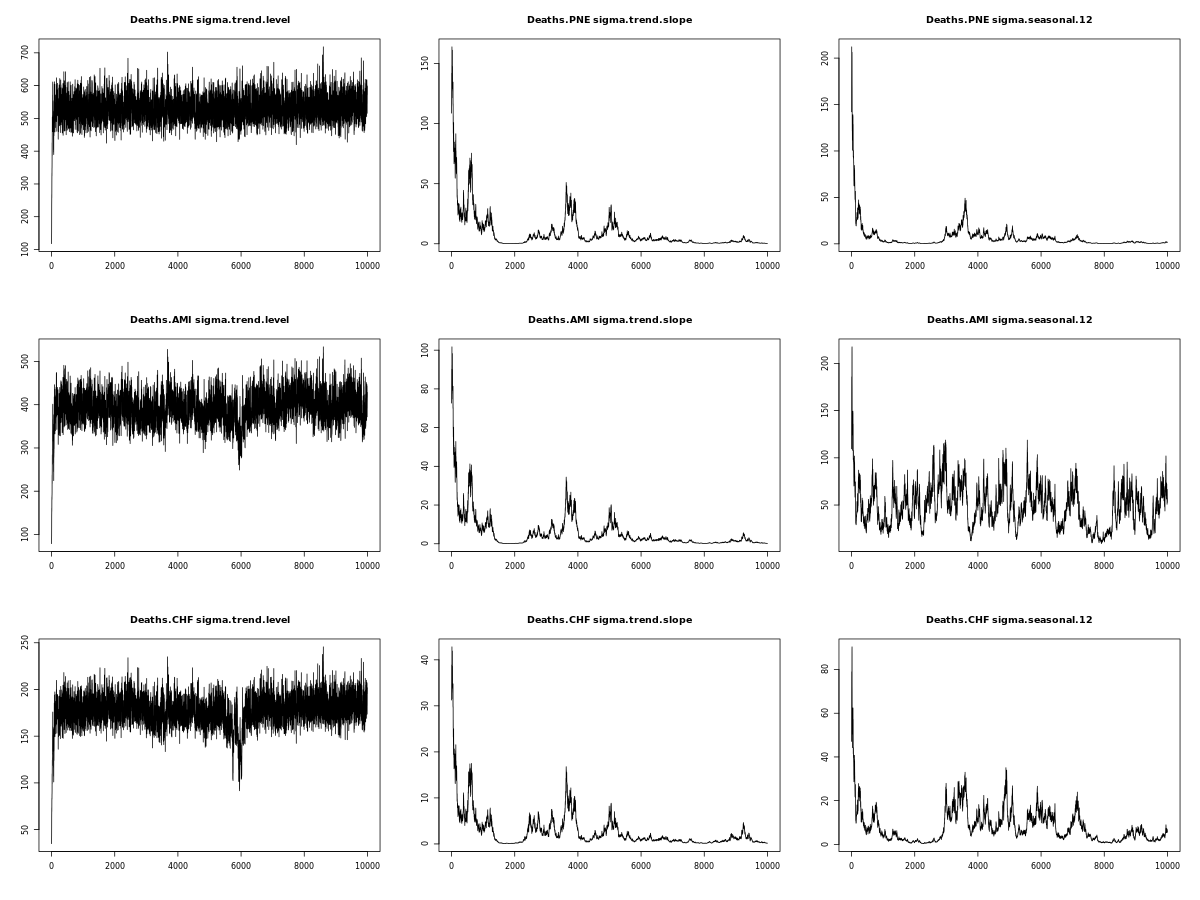}

\vspace{20pt}

\caption{Trace plots for the alternative \textbf{local level and seasonal model without regressors} estimated in the pre-intervention period on the death counts from each condition.}
\includegraphics[scale=0.3]{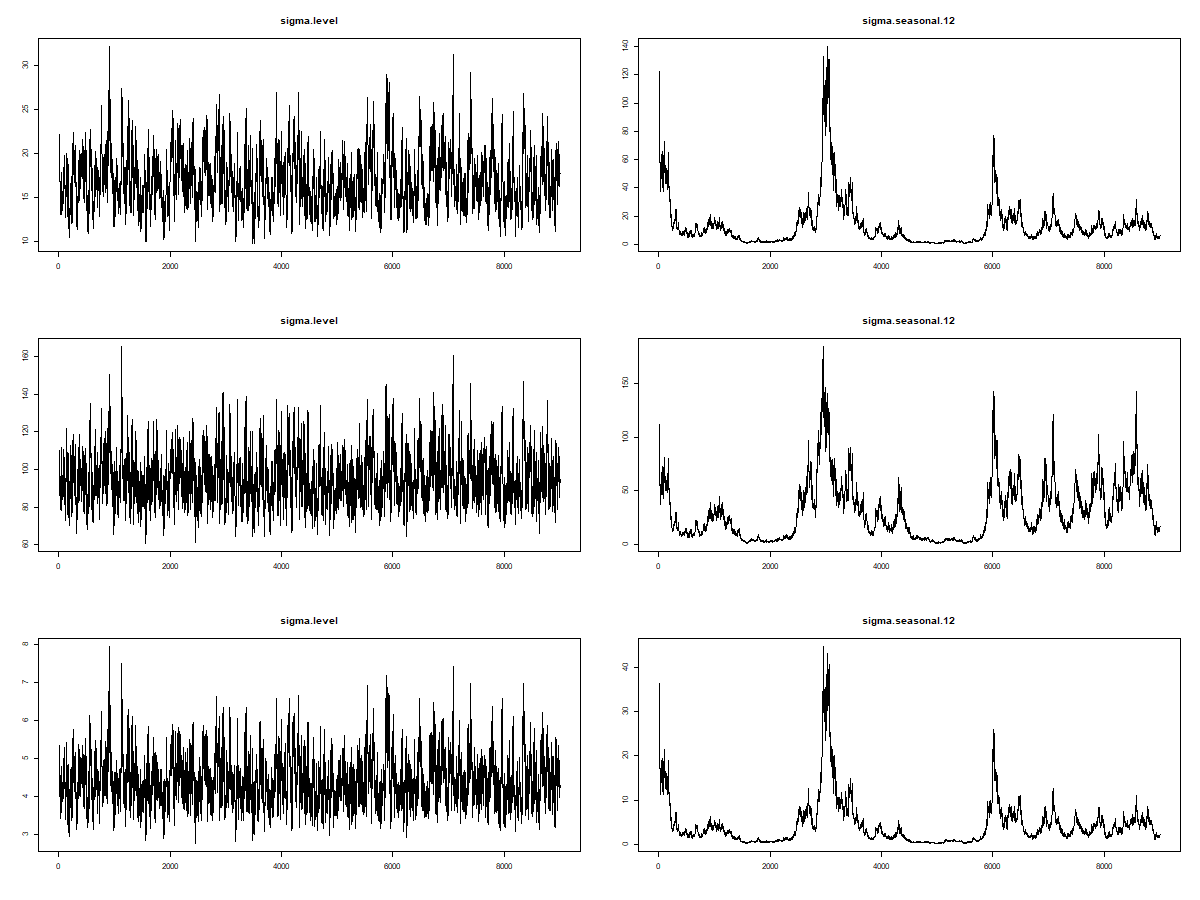}
\label{fig:trace_llevel_nocov}
\end{figure}

\begin{figure}[p]
\centering
\caption{Trace plots for the alternative \textbf{local level and seasonal model with regressors} estimated in the pre-intervention period on the death counts from each condition.}
\includegraphics[scale=0.3]{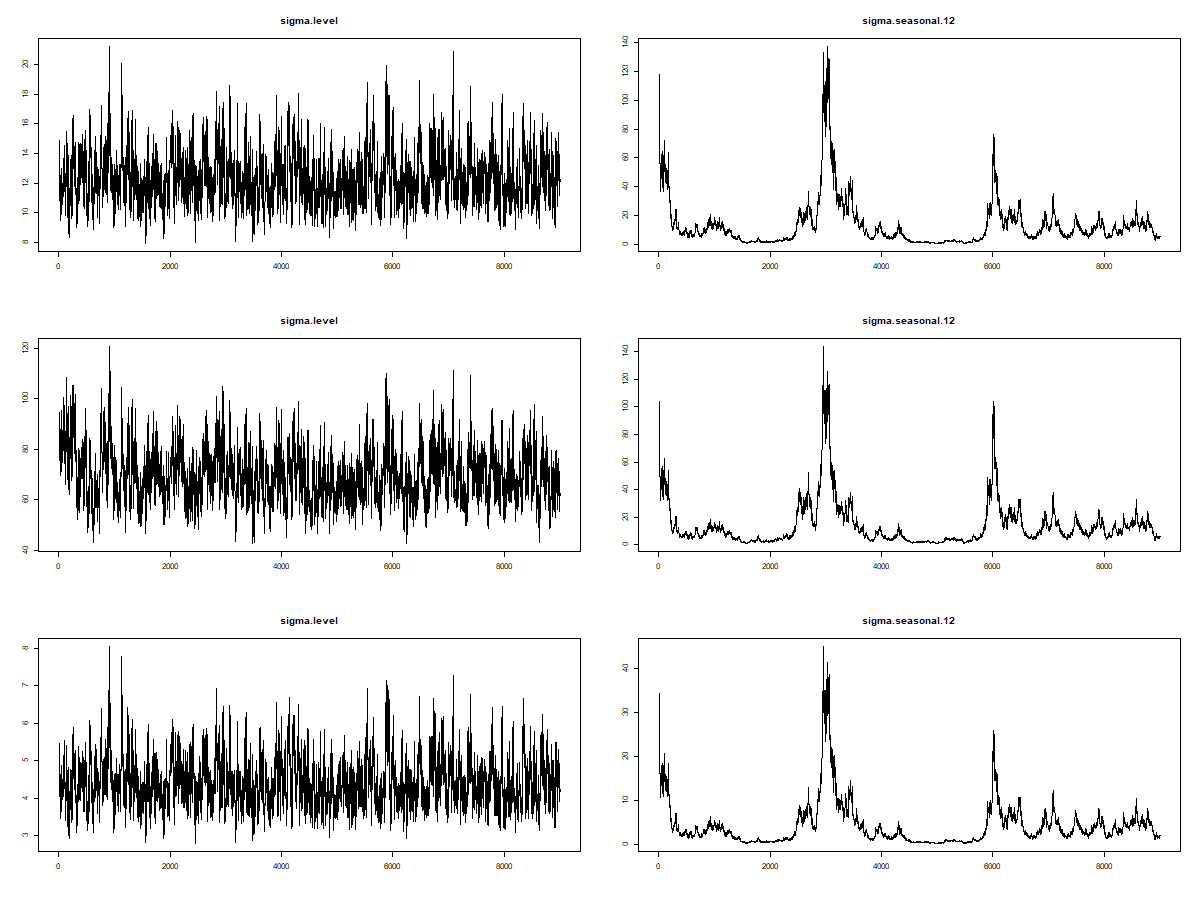}
\label{fig:trace_llevel}

\vspace{20pt}

\caption{Trace plots for a \textbf{local linear trend and seasonal model excluding PM$_{2.5}$} from the set of regressors.}
\includegraphics[scale=0.33]{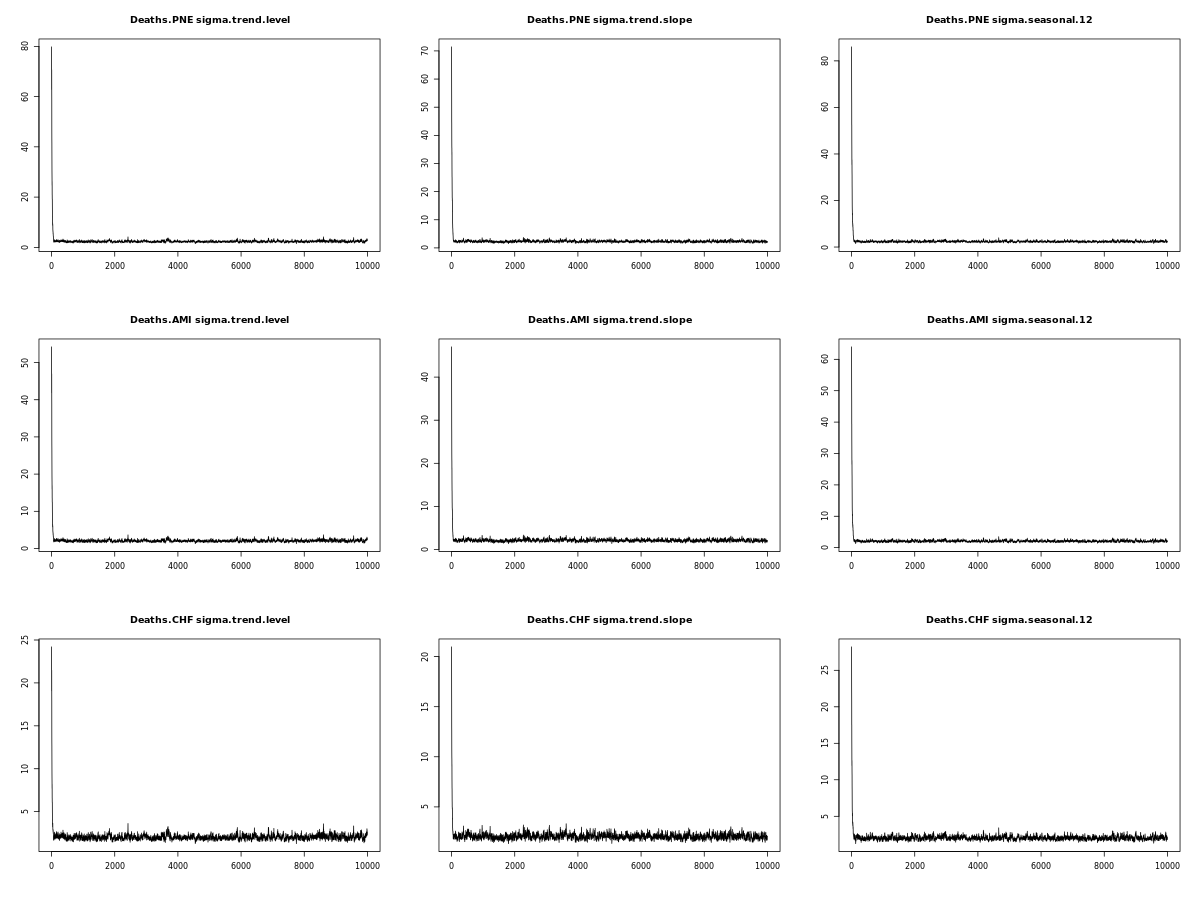}
\label{fig:trace_no_pm25}
\end{figure}

\begin{figure}[!t]
\centering
\caption{Trace plots for a \textbf{local linear trend and seasonal model having as regressors the weather covariates only}.}
\includegraphics[scale=0.33]{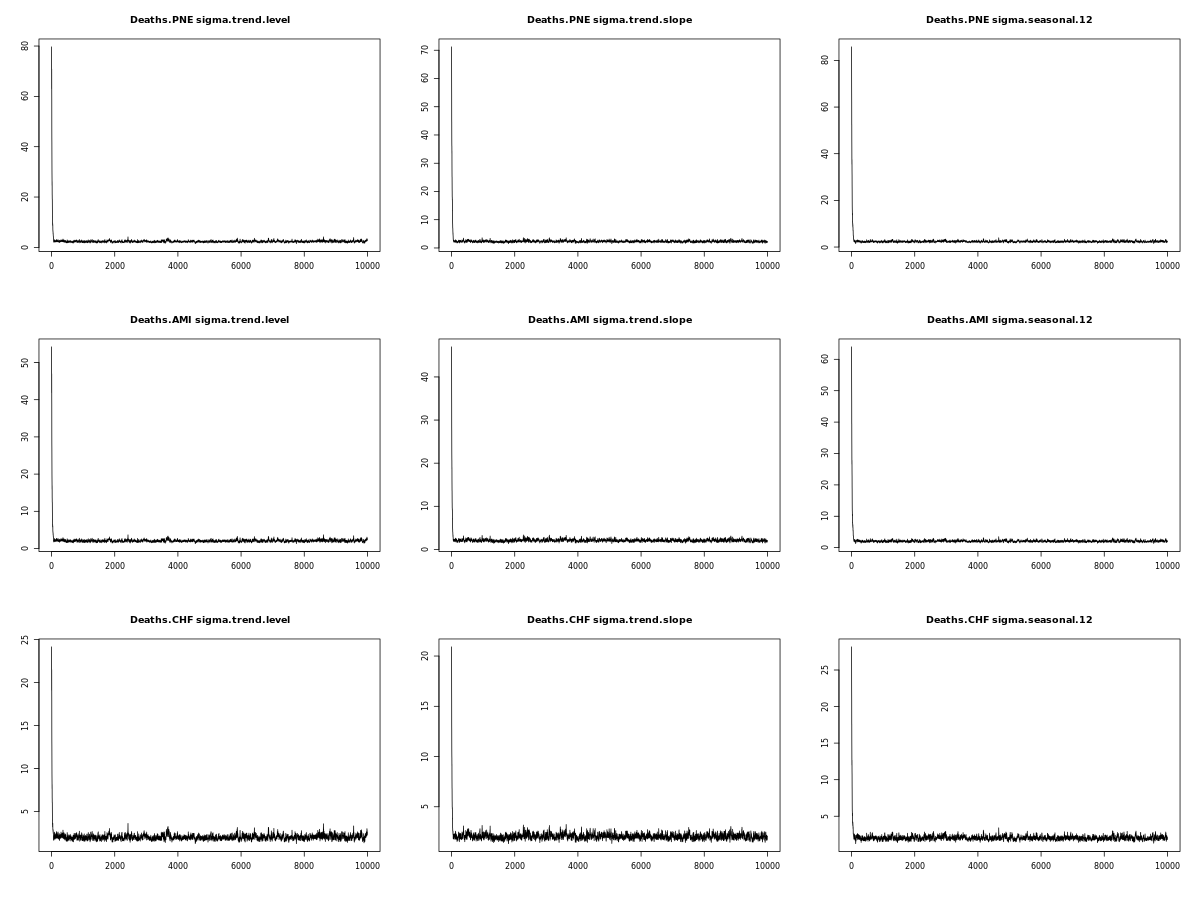}
\label{fig:trace_only_weather}
\end{figure}

We investigated trace plots of identifiable parameters to visually inspect the convergence of the Markov chain to the stationary distribution.
Figures \ref{fig:trace_llinear} -- \ref{fig:trace_only_weather} show the trace plots of the alternative models that we tested in the same order as in \cref{subsec:ppc}.

We notice that the local level models have convergency issues in the variance of the seasonal component  (Figures \ref{fig:trace_llevel_nocov} and \ref{fig:trace_llevel}). Instead, the selected local linear and seasonal model seems to converge to the stationary distribution. The models for which the MCMC showed lack of convergence were {\it not} used to evaluate the causal effect.

\vfill

\vfill

\clearpage

\restoregeometry
\restoregeometry

\subsection{Posterior inclusion probabilities}

The prior distribution on the regression coefficients $\beta$ is a spike-and-slab prior defined after a data augmentation step where we add a selection vector whose components can either be 0 (the corresponding regressor is deleted from the model) or 1 (the corresponding regressor is included in the model). At each MCMC iteration, a stochastic search variable selection is performed: the Gibbs sampling algorithm inspects all possible models by changing the selection vector component-wise, so that, at each iteration, only the most likely model is retained. By averaging across all selected models (one for each MCMC iteration) we obtain the posterior inclusion probabilities for each regressor. These are useful to understand whether the covariates that we choose are related to the outcome and which ones retain most of the information to improve the prediction of the outcome in the pre-intervention period.

\begin{figure}[H]
\centering
\caption{Posterior inclusion probabilities of the covariates included in the selected local linear and seasonal models estimated in the pre-intervention period on the death counts from each condition at the national level.}
\label{fig:inclusion_probs}
\includegraphics[scale=0.58, trim = 0 0 800 0, clip]{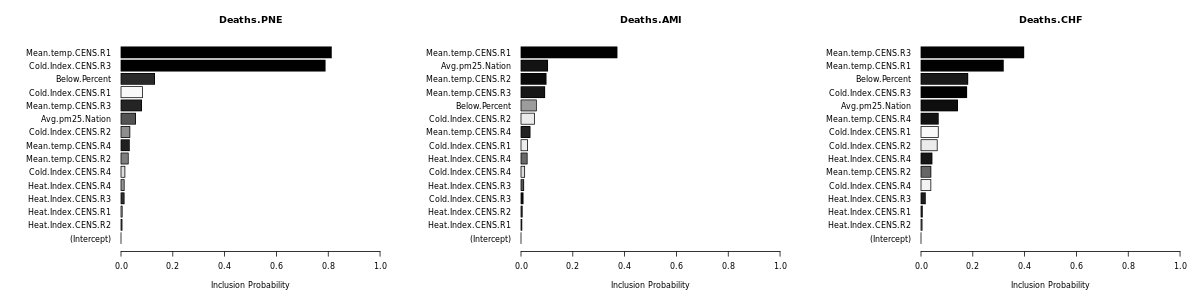}
\includegraphics[scale=0.58, trim = 400 0 400 0, clip]{inclusion_probs_national.png} \\
\includegraphics[scale=0.58, trim = 800 0 0 0, clip]{inclusion_probs_national.png}
\end{figure}

\begin{figure}[p]
\centering
\caption{Posterior inclusion probabilities of the covariates included in the selected local linear and seasonal models estimated in the pre-intervention period on the death counts from each condition, by region.}
\includegraphics[scale = 0.45]{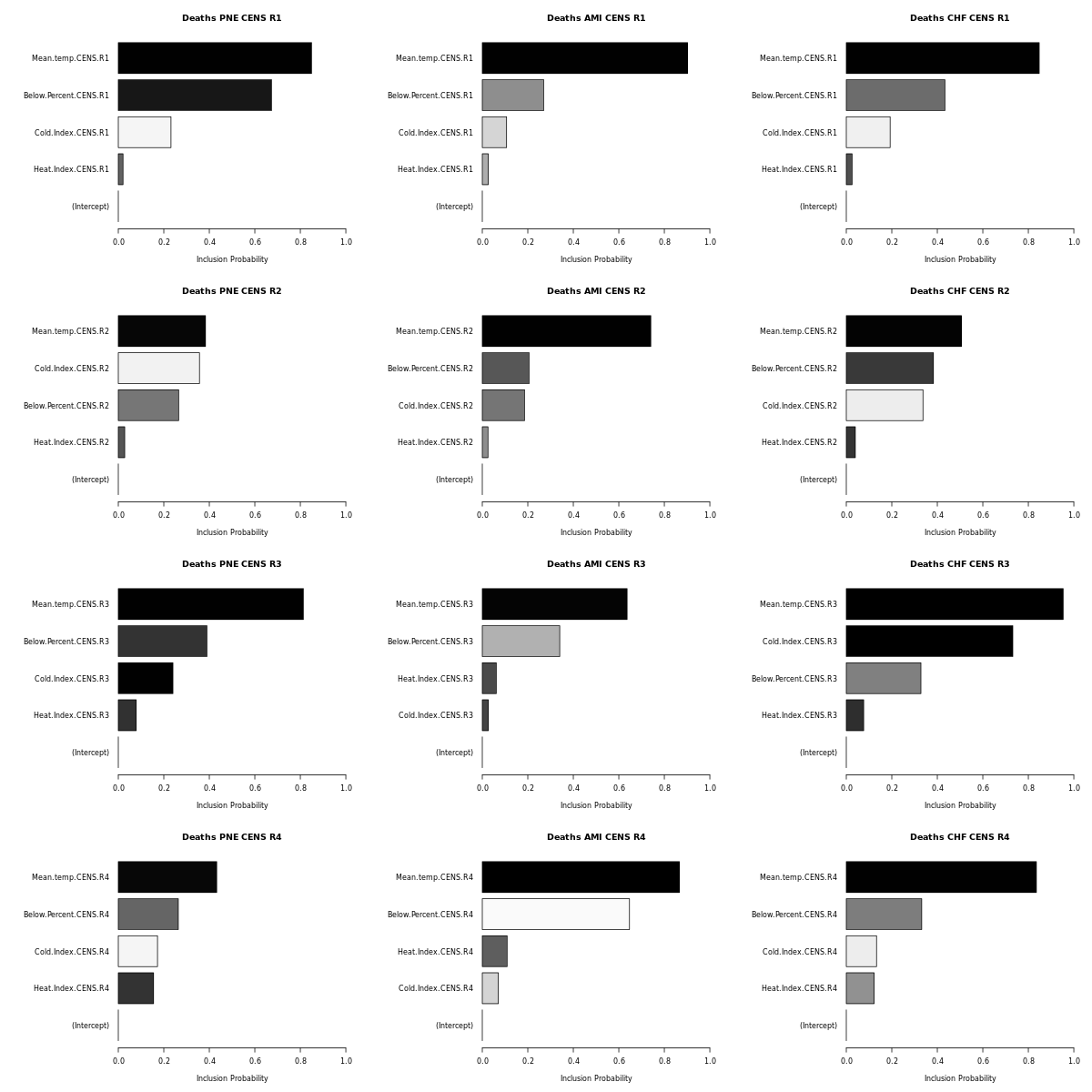}
\label{fig:inclusion_probs_regional}
\end{figure}

\cref{fig:inclusion_probs} shows the results for the national analyses.
We notice that temperature information seems to play a crucial role in modeling deaths from all conditions at the national level, followed by poverty. This evidence is supported by the regional analyses as well, results of which are shown in \cref{fig:inclusion_probs_regional}. Interestingly, from the regional breakdown we observe that poverty is linked to death counts especially in the South and West regions (its posterior inclusion probability in the model for CHF is 60\% in the South and exceeds 80\% in the West). 

\vfill

\vfill

\clearpage

\newgeometry{top = 3cm, bottom = 3cm, left = 1cm, right = 1cm}

\restoregeometry
\restoregeometry

\section{Additional plots of the estimated causal effects}

\begin{figure}[h!]
\centering
\caption{Cumulative effect of the HRRP on mortality from pneumonia, AMI and CHF for the remaining regions at the intermediate time horizon ending in December 2015.}
\label{fig:regional_impacts}
\includegraphics[scale=0.45]{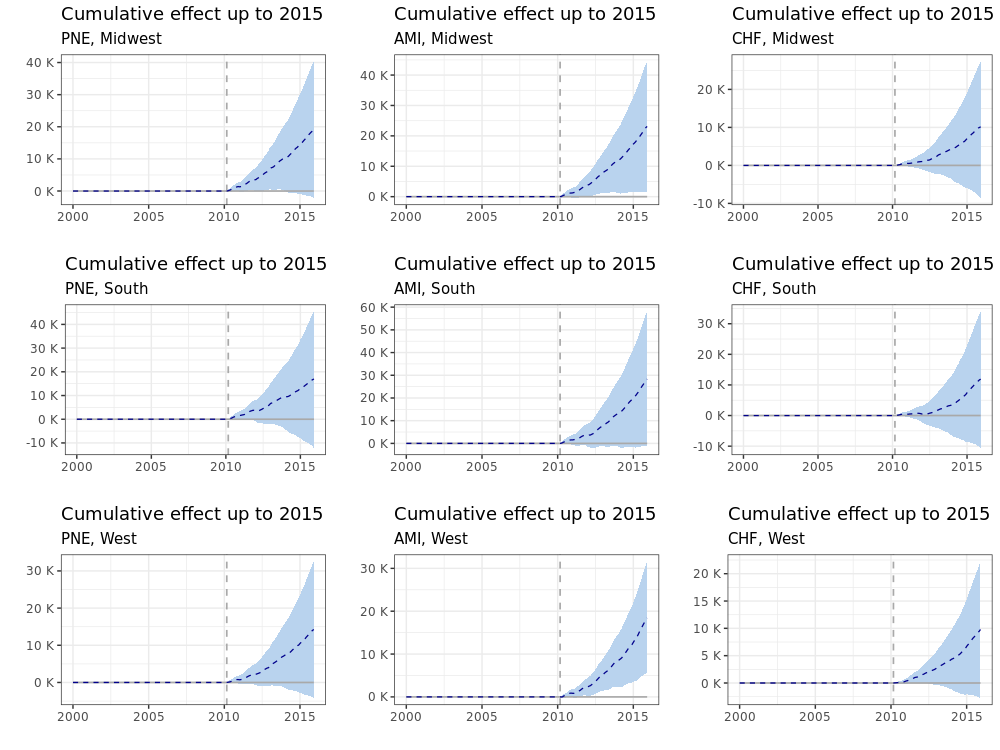}
\end{figure}

\begin{figure}[p]
\centering
\caption{Comparison of the observed series with the predicted outcome from the selected local linear and seasonal model at the intermediate time horizon ending in December 2015, by region.}
\label{fig:regional_predicted}
\includegraphics[width = \textwidth]{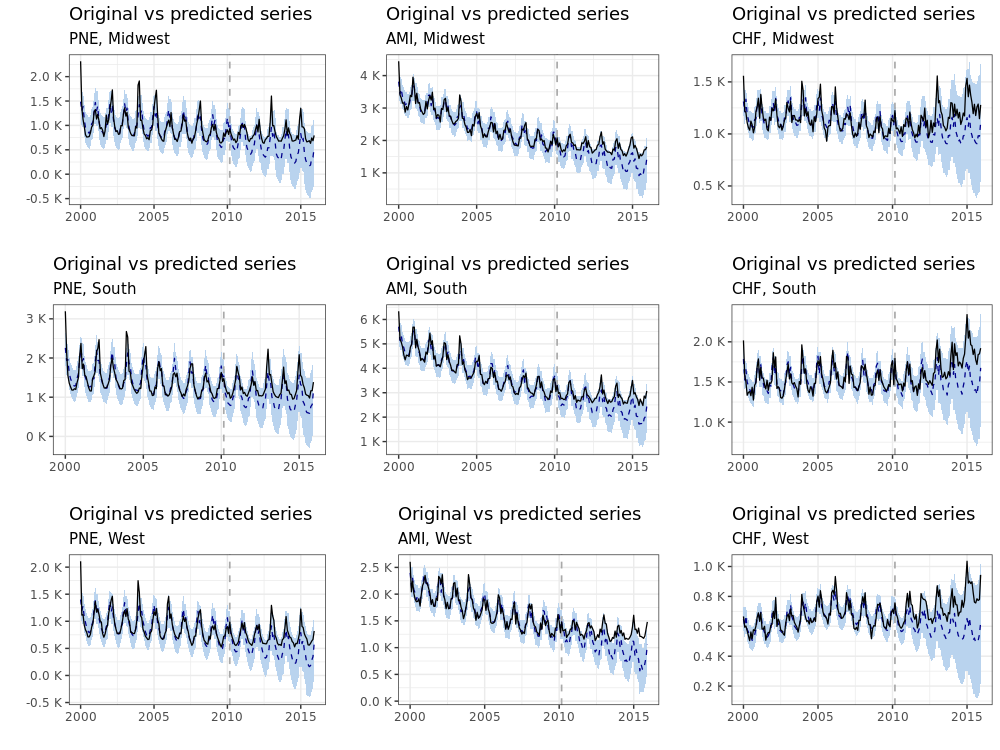}
\end{figure}

We now display additional plots of the estimated causal effects. Figure \ref{fig:regional_impacts} provides a visual overview of the mid-term cumulative impacts generated by the HRRP on mortality for the remaining regions (Midwest, South, West). In addition, Figure \ref{fig:regional_predicted} shows the comparison between the observed series and the predicted outcome based on the model fitted to the pre-intervention data. We notice that the fitted values before the intervention are very close to the observed values, which is a further indicator of model adequacy. Therefore, based on Assumption \ref{ass:model}, we can attribute the deviation from the observed series in the post-intervention period to the HRRP (the deviation is also visible in the Figure).

\vfill

\vfill
\clearpage

\newgeometry{top = 3cm, bottom = 3cm, left = 1.3cm, right = 1.3cm}

\restoregeometry
\restoregeometry

\clearpage

\section{Robustness checks}
\label{supp_sec:robustness_checks}

Conscious of the policy implication of our findings, we run several checks to see if the results are robust to different model specifications. 
We run additional analyses with different model specifications and using alternative methods that are commonly used in causal inference literature. The next sections summarize the main results.

\subsection{Alternative choice of predictors}

First of all, we performed a sensitivity analysis aimed at assessing if and how much the estimates depend on different choices of predictors. The results are reported in Section \ref{subsec:program_evaluation} of the main paper and show that the empirical findings for pneumonia, AMI and CHF are consistent across different sets of regressors. Diagnostic tests were shown in \cref{supp_sec:model_fit}.

\subsection{Falsification tests}
\label{subsec:falsification}

\begin{figure}[!t]
\centering
\caption{Evolution over time of the control outcomes at the US level. The dashed vertical bar indicates the date when the HRRP was introduced.}
\label{fig:fals_outcomes}
\includegraphics[width=\textwidth]{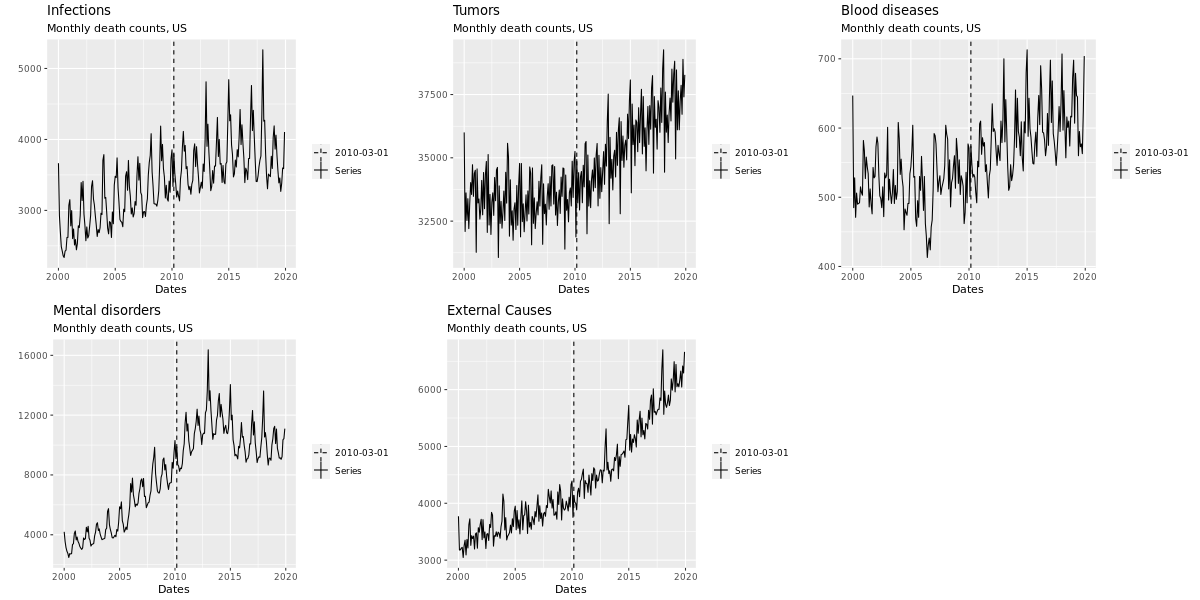}
\end{figure}

We run falsification tests by repeating the analyses on control outcomes that, in principle, should be unaffected by the HRRP incentive scheme. Figure \ref{fig:fals_outcomes} displays the evolution of the control outcomes over the analysis period. Under the usual model specification (local linear trend and seasonal model estimated on count data) no causal effect is found on control outcomes (see Table \ref{tab:res_falsification} in the main paper). We also considered whether the control conditions indicated an anticipatory effect by fictionally setting the treatment initiation period as three months earlier and using our method to evaluate the effect in the three month time period that follows. In \cref{tab:anticipation}, we see that none of the control conditions (or the target conditions) shows an anticipatory effect.

\clearpage

\begin{table}[!t]
\centering
\caption{Anticipation effects of the HRRP on the falsification outcomes up to 3 months before the intervention.}
\label{tab:anticipation}
\small
\begin{tabular}{lrrrr}
  \hline
& C$\Delta$ & 2.5\% & 97.5\% & \\ 
  \hline
Infections       & -685  & -1,362 & 52 & \\ 
Tumors           & 952   & -1,113 & 2,881 & \\ 
Blood diseases   & -55   & -210  & 99 &  \\ 
Mental disorders & 1,143  & -1,655 & 4,219 \\ 
External causes  & -11   & -578  & 558 \\
\hline
Pneumonia & -1,323 & -4,112 & 1,329\\ 
AMI & -1,344 & -3997 & 1,338 \\ 
CHF & 579  & -390  & 1,451 \\ 
   \hline
\end{tabular}
\end{table}

\clearpage

\restoregeometry
\restoregeometry

\subsection{Interrupted time series approach}
\label{subsec:its}

\begin{table}[!b]
\centering
\caption{Estimated post-intervention slope change in a segmented regression for pneumonia, AMI and CHF at three different time horizons. Significant results at the $5$\% level are shown in bold.}
\label{tab:its_segmented}
\scalebox{0.9}{
\begin{tabular}{lccccccccccc}
  \hline \hline
  & \multicolumn{3}{c}{PNE} & & \multicolumn{3}{c}{AMI} & & \multicolumn{3}{c}{CHF} \\
 \cline{2-12}
& $\beta_3$ & Std. Error & t stat & & $\beta_3$ & Std. Error & t stat & & $\beta_3$ & Std. Error & t stat \\
  \hline
Short-term & 6.94  & 18.80 & 0.37 & & 29.02 & 15.29 & 1.90  & & -2.80 & 6.51 & -0.43 \\ 
Mid-term   & \bf 12.24 & \bf 5.57  & \bf 2.20 & & \bf 43.49 & \bf 4.52  & \bf 9.62  & & \bf 12.48 & \bf 2.11 & \bf 5.92 \\ 
Long-term  & \bf 14.09 & \bf 4.86  & \bf 2.90 & & \bf 46.39 & \bf 3.96  & \bf 11.71 & & \bf 12.32 & \bf 2.06 & \bf 5.98 \\ 
   \hline \hline
\end{tabular}
}
\end{table}

For comparison purposes, we repeated the analysis using the Interrupted Time Series (ITS) approach. The usual ITS design applies when there are policy changes affecting all units simultaneously. Such design would then be a very natural choice in our setting, since the HRRP has the characteristics of an extensive policy affecting all hospitals and patients at the same time. 
Segmented regression is the simplest form of ITS analysis and involves estimating a model as the following \citep{Wagner:Soumerai:Zhang:2002, Bernal:Cummins:Gasparrini:2017},
\begin{equation}
\label{eqn:segmented}
    Y_t = \beta_0 + \beta_1 t + \beta_2 D_t + \beta_3 tD_t + \epsilon_t 
\end{equation}
where: $t$ is the time since the start of the study; $D_t$ is an indicator variable coded $0$ before the intervention and $0$ after the intervention; $t D_t$ is the interaction between the indicator variable and the time and counts the time points after the intervention (it is $0$ before the policy introduction); $\epsilon_t \sim i.i.d N(0, \sigma^2)$ denote the error terms, which are typically assumed to be uncorrelated, homoschedastic and independent. Thus, under such formulation, $\beta_0$ and $\beta_1$ would capture, respectively, the baseline level and trend; $\beta_2$ estimates the level change after the intervention and $\beta_3$ estimates the change in trend. Model equation (\ref{eqn:segmented}) can then be adapted to the characteristics of the data. For example, in our case we also added covariates reflecting the presence of possible confounders linked to the treatment and the outcome and seasonality effects (captured by the inclusion of dummy variables). The results are reported in Table \ref{tab:its_segmented}. For brevity reasons, the table only shows the coefficient estimates for $\beta_3$, since the estimates for the level change, $\beta_2$, was never significant at the $5$\% level.\footnote{Full results are available upon request.} Results seem to support the idea that mortality trend increased after the HRRP both in the mid-term and in the long-term for all conditions. We also used these models to predict the outcome had the intervention not taken place by setting the corresponding treatment indicator to 0. We used these predicted outcomes to acquire estimates of $C\Delta_K$ for the short-, mid-, and long-term. These results are shown in \cref{tab:its_segmented_effects}, though it is not clear how one could acquire valid confidence intervals for these quantities based on this model. We see that the magnitude of the effect estimates based on the ITS analyses is comparable to the magnitude of the results based on our method.

\begin{table}[!t]
\centering
\caption{Cumulative effect estimates for pneumonia, AMI and CHF at three different time horizons based on the ITS analysis.}
\label{tab:its_segmented_effects}
\scalebox{0.9}{
\begin{tabular}{lccc}
  \hline \hline
  & pneumonia & {AMI} & {CHF} \\
  \hline
Short-term &  2,738 & 15,445 & 747 \\ 
Mid-term   & 36,199 & 126,074 & 27,356 \\ 
Long-term  & 101,941 & 347,064 & 77,266 \\ 
   \hline \hline
\end{tabular}
}
\end{table}

However, segmented regression models do not typically account for autocorrelation. To incorporate that, we should allow the error term to follow an ARIMA process, as it is recommended in \citet{Schaffer:Dobbins:Pearson:2021}. In addition, seasonality can also be formally handled by differencing, i.e., calculating the difference between adjacent observations (for a detailed review of ARIMA models, see \citet{Brockwell:2009}). By applying this less trivial ITS approach to our data we obtain a significant increase in mortality trends for pneumonia and CHF. The results are reported in Table \ref{tab:its_arima}. 

\begin{table}[!b]
\centering
\caption{Estimated post-intervention slope change in an ITS regression with ARIMA errors for pneumonia, AMI and CHF at three different time horizons. Significant results at the $5$\% level are shown in bold.}
\label{tab:its_arima}
\scalebox{0.9}{
\begin{tabular}{lccccccccccc}
  \hline \hline
  & \multicolumn{3}{c}{PNE} & & \multicolumn{3}{c}{AMI} & & \multicolumn{3}{c}{CHF} \\
 \cline{2-12}
& $\beta_3$ & Std. Error & t stat & & $\beta_3$ & Std. Error & t stat & & $\beta_3$ & Std. Error & t stat \\
  \hline
Short-term & 3.68  & 16.41 & 0.22 & & 29.41 & 17.25 & 1.70 & & 3.35 & 7.01  & 0.48 \\ 
Mid-term   & \bf 12.97 & \bf 6.28  & \bf 2.06 & & 25.99 & 14.18 & 1.83 & & \bf 12.87 & \bf 2.42 & \bf 5.32 \\ 
Long-term  & \bf 13.34 & \bf 5.95  & \bf 2.24 & & 31.55 & 17.26 & 1.83 & & \bf 10.94 & \bf 2.30 & \bf 4.75 \\
   \hline \hline
\end{tabular}
}
\end{table}

Both segmented regression models and ITS models with ARIMA errors are estimated on the full pre-intervention and post-intervention data, which implies that one must postulate a specific structure on the effect, e.g., a pulse, a step change, a transient change. For example, results above have been obtained under the assumption that the HRRP produced a ``ramp'' effect on the outcome, i.e., a change in slope that occurs after the intervention \citep{Schaffer:Dobbins:Pearson:2021}. However, if this assumption is wrong, the estimated effect will necessarily be biased. In addition, to produce effect estimates at different point in time, model estimation has to be repeated multiple times adjusting the length of the time series, which is time consuming and sub-optimal (the estimated coefficient is an average over the period).\footnote{For a detailed discussion on the pros and cons of common ITS approaches see \citet{Menchetti:Cipollini:Mealli:2021}, where a simulation study also shows that causal effect estimators are biased when the assumed structure of the effect is wrong.} A few recent approaches overcome this limitation \citep{Miratrix2019simulating, Menchetti:Cipollini:Mealli:2021}. In particular, \citet{Miratrix2019simulating} propose a straightforward generalization of ITS designs by performing model estimation up to the intervention time point and then simulate different outcome trajectories from the theoretical distributions of estimated model parameters. In this way, no structure is imposed on the causal effect.

We therefore use the ITS approach proposed by \citet{Miratrix2019simulating} to estimate the causal effect of the HRRP on mortality from pneumonia, AMI and CHF. The results are reported in Table \ref{tab:its} and are analogous to those reported in the paper obtained with the Bayesian structural time series approach. All the estimates have been obtained with the \texttt{simITS} R package \citep{simITS}. 

\begin{table}[!b]
\centering
\caption{Cumulative causal effect and $95 \%$ credibility intervals of the HRRP on mortality from pneumonia, AMI and CHF at the national level estimated with the ITS approach of \citet{Miratrix2019simulating}. In this table, C$\Delta$ denotes the cumulative impact at the end of three different time horizons (short, mid, and long-term). Results with 95\% CIs that do not include 0 are shown in bold.}
\label{tab:its}
\scalebox{0.9}{
\begin{tabular}{lccccccccccc}
  \hline \hline
  & \multicolumn{3}{c}{PNE} & & \multicolumn{3}{c}{AMI} & & \multicolumn{3}{c}{CHF} \\
 \cline{2-12}
& C$\Delta$ & 2.5\% & 97.5\% & & C$\Delta$ & 2.5\% & 97.5\% & & C$\Delta$ & 2.5\% & 97.5\% \\
  \hline
Short-term & -4,929 & -30,035  & 18,339  & & 9,329   & -16,313 & 32,127  & & -3,956 & -13,891 & 5,096 \\ 
Mid-term   & -900   & -83,367  & 75,484  & & \bf 94,515  & \bf 17,271  & \bf 168,923 & & 14,311 & -16,547 & 44,342 \\ 
Long-term  & 1,901  & -166,226 & 157,712 & & \bf 261,164 & \bf 101,582 & \bf 414,572 & & 51,096 & -13,740 & 109,168 \\ 
   \hline \hline
\end{tabular}
}
\end{table}

\subsection{Difference-in-Differences}
\label{subsec:did_synth}

A widely adopted method to evaluate the impact of an intervention on time series data is Difference-in-Difference (DiD) (see e.g., \citet{Card:Krueger:1993, Meyer:Viscusi:Durbin:1995, Angrist:Pischke:2008}). In its simplest formulation, this method requires to observe a treated and a control group at a single point in time before and after the intervention; the effect is then estimated by contrasting the change in the average outcome for the treated group with that of the control group. 
However, DiD often relies on the assumption that, in the absence of treatment, the outcomes of the treated and control units would have followed parallel paths. Therefore, before using DiD for the purpose of a causal analysis, it is generally recommended to check whether pre-intervention trends of the treated and control outcomes are parallel \citep{Ryan:Burgess:Dimick:2015, ONeill:Kreif:Grieve:Sutton:Sekhon:2016}.

\begin{figure}[p]
\centering
\caption{Pre-intervention trends of pneumonia, AMI and CHF compared to that of possible control conditions. To ease comparison by correcting for different scales, death counts are scaled by their starting values in January 2000 (base month). Blue is used for the target condition, and black for the average of the control conditions.}
    \label{fig:did_pre_trends}
{\large \bf Monthly death counts in the pre-intervention period} \\[20pt]
\subfloat[PNE]{\includegraphics[width = 0.7\textwidth,trim=0 0 0 40, clip]{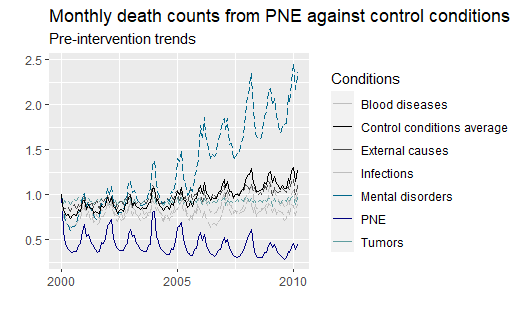}} \\[-15pt]
\subfloat[AMI]{\includegraphics[width = 0.7\textwidth,trim=0 0 0 40, clip]{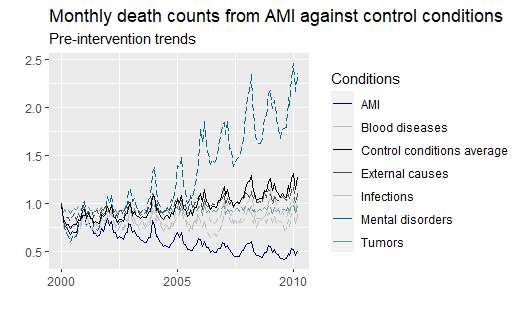}} \\[-15pt]
\subfloat[CHF]{\includegraphics[width = 0.7\textwidth,trim=0 0 0 40, clip]{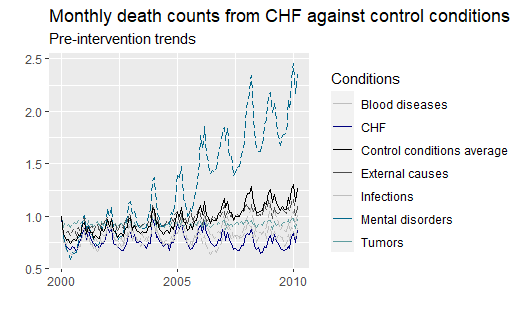}}
\end{figure}

In our case, we attempted to use the DiD approach taking as control conditions the control outcomes analysed in Section \ref{subsec:falsification}, since we found that mortality from such conditions was unaffected by the HRRP. We therefore checked the validity of the parallel trend assumption by plotting the pre-intervention mortality trends from our outcomes and the control outcomes. As we can see from Figure \ref{fig:did_pre_trends}, the pre-intervention trends of the control conditions are not parallel to those exhibited by pneumonia, AMI and CHF. For this reason, we do not recommend to adopt DiD in our specific setting.

\subsection{Synthetic control methodology}

An alternative approach that does not rely on the parallel trend assumption is the synthetic control method \citep{Abadie:Gardeazabal:2003, Abadie:Diamond:Hainmueller:2010, Abadie:Diamond:Hainmueller:2015}, which, instead, creates a synthetic time series by taking a weighted average of control series in a ``donor pool'' that are unaffected by the intervention. The weights are found by minimizing the distance between pre-treatment outcomes and the synthetic time series. The weights are usually restricted to be non-negative (non-negativity) and to sum up to 1 (adding-up) but several generalizations have been recently introduced. Indeed, relaxing these restrictions may be useful when the outcome of interest is an outlier with respect to the other units and in case of negative correlations between the outcome and the control series, thereby increasing prediction accuracy in the post-intervention period \citep{Doudchenko:Imbens:2016}. 

\begin{figure}[!b]
    \centering
    \caption{Estimated pointwise causal effects of HRRP under the augmented synthetic control method.}
    \label{fig:aug_synth}
    \includegraphics[scale = 0.55]{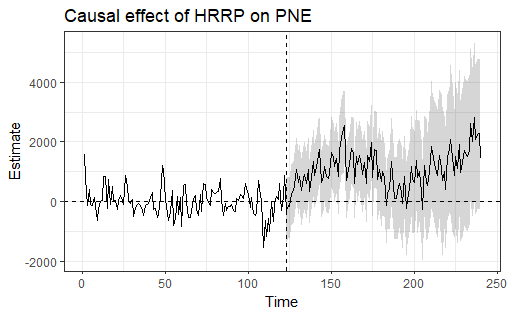}
    \includegraphics[scale = 0.55]{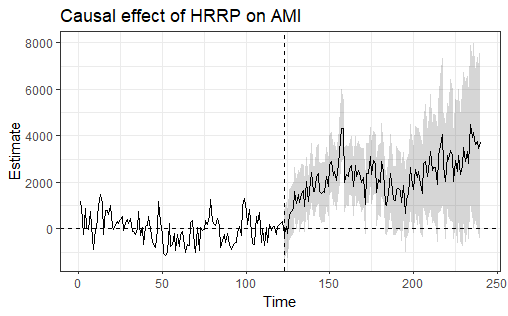}
    \includegraphics[scale = 0.55]{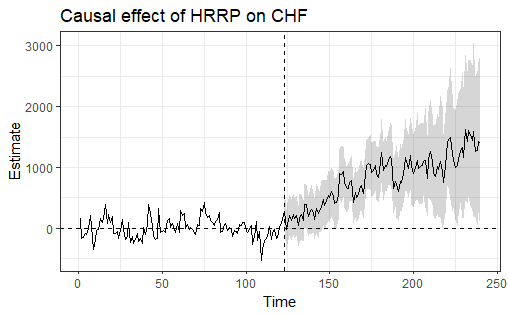}
\end{figure}

We therefore adopt the augmented synthetic control method proposed in \citet{Benmichael:Feller:Rothstein:2021}, which relaxes both the adding-up and the non-negativity restrictions and augments the common synthetic control approach with an outcome model that allows to remove the bias due to an imperfect pre-treatment fit. This method also permits to include auxiliary covariates by directly including them into the balance objective for the synthetic control method, as it is done for the controls. We use time series from conditions that are not expected to be impacted by the intervention as the control pool, similarly to \cite{gaughan2019paying}. All the estimates have been obtained using the \texttt{augsynth} R package\footnote{This package is released on GitHub and it is available at \url{https://github.com/ebenmichael/augsynth}.}. Figure \ref{fig:aug_synth} plots the estimated causal effect of the HRRP on each condition. The estimated cumulative effects $C\Delta_k$ in the short, medium and long run are reported in Table \ref{tab:aug_synth} and have been obtained by adding up the impact estimates at each point in time after the intervention. Although it remains unclear how to obtain valid confidence intervals for these estimates, we can notice, once again, that their magnitude is aligned with the results reported in Table \ref{tab:res} under the selected Bayesian structural time series approach, though again slightly larger.

\begin{table}[!t]
\centering
\caption{Cumulative effect estimates for pneumonia, AMI and CHF at three different time horizons based on the augmented synthetic control analysis.}
\label{tab:aug_synth}
\begin{tabular}{rrrr}
  \hline \hline
 & pneumonia & AMI & CHF \\ 
  \hline 
short-run & 15,286 & 27,972 & 4,626 \\ 
mid-run   & 68,349 & 134,961 & 41,120 \\ 
long-run  & 126,547 & 265,566 & 96,580 \\ 
   \hline \hline
\end{tabular}
\end{table}

\subsection{Alternative model specifications based on transformations of our outcome variable}

Referring to the main model with local linear trend, seasonality and predictors as ``Model 1'', we also tested alternative model specifications, namely:
\begin{enumerate}
\item Gaussian Bayesian structural time series model estimated on the logarithmic transformation of the monthly death counts (Model 2);
\item Gaussian Bayesian structural time series model estimated on the logit transformation of the monthly death rate from each alternative condition, i.e., monthly death counts divided by the over 65 population (Model 3).
\end{enumerate}

\begin{figure}[p]
\centering
\caption{Posterior predictive checks of control outcomes under Model 1.}
\label{fig:ppc_m1}
\includegraphics[width=\textwidth]{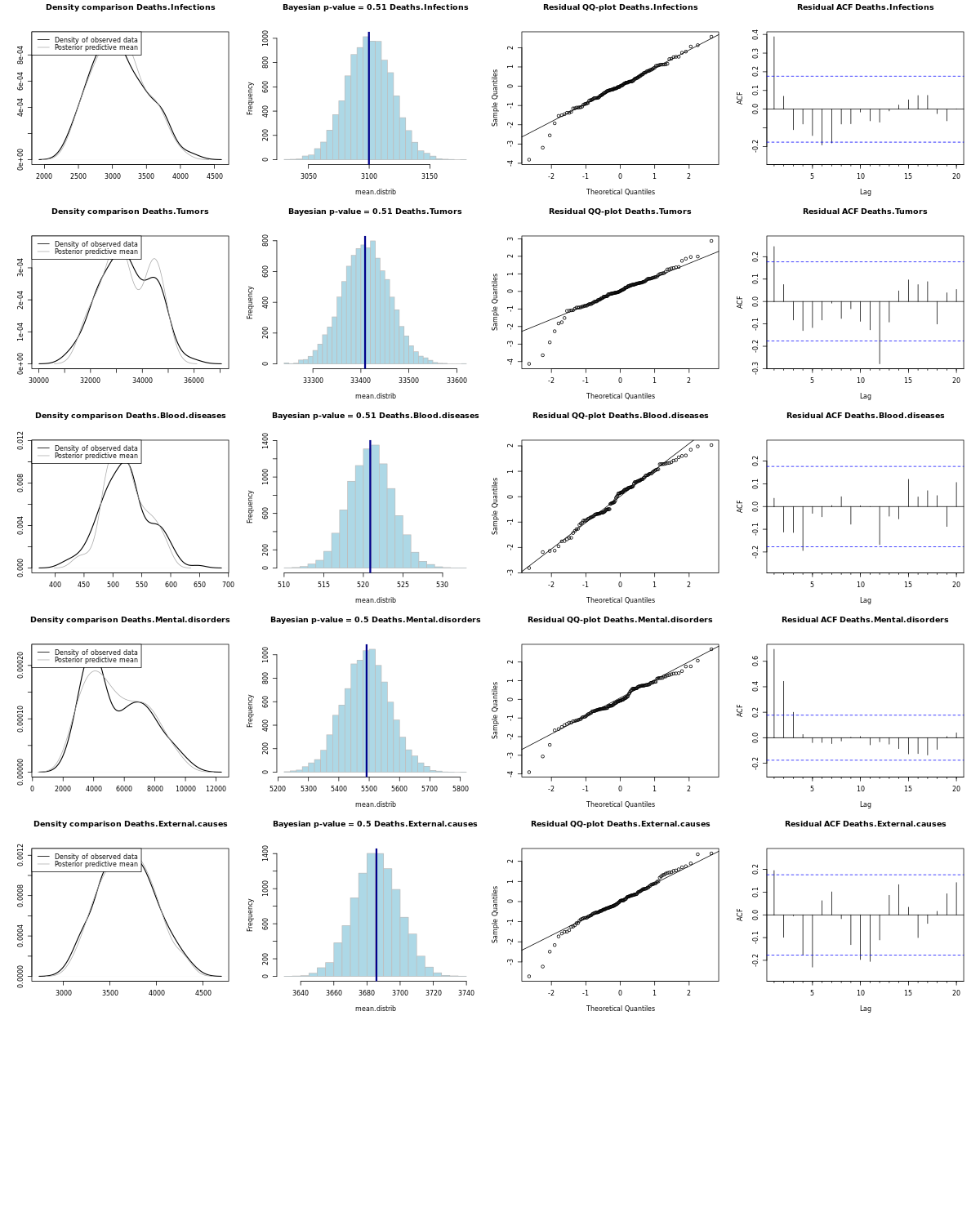}
\end{figure}

\begin{figure}[p]
\centering
\caption{Posterior predictive checks for pneumonia, AMI and CHF under Model 2.}
\label{fig:ppc_m2_true}
\includegraphics[width=\textwidth]{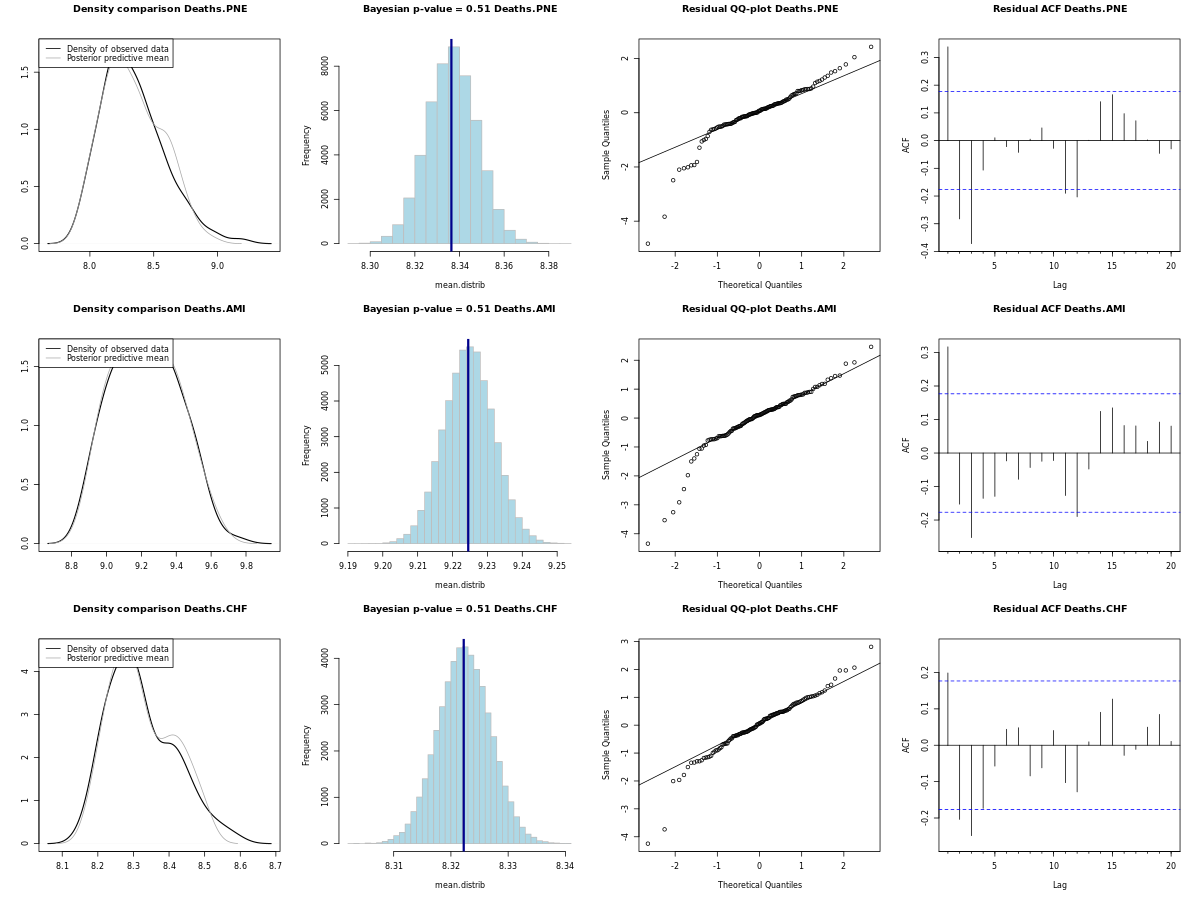}
\end{figure}

\begin{figure}[p]
\centering
\caption{Posterior predictive checks of control outcomes under Model 2.}
\label{fig:ppc_m2}
\includegraphics[width=\textwidth]{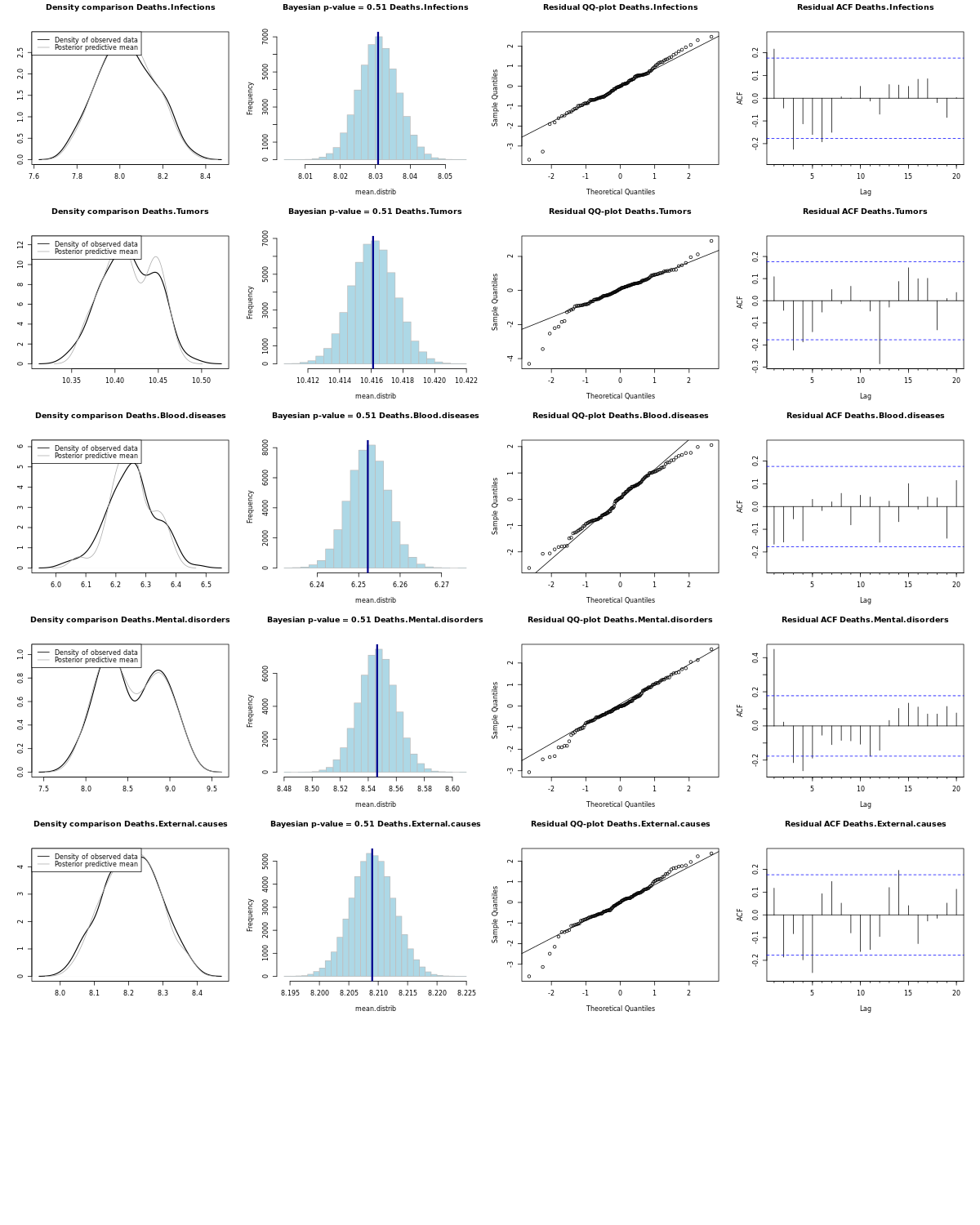}
\end{figure}

\begin{figure}[p]
\centering
\caption{Posterior predictive checks for pneumonia, AMI and CHF under Model 3.}
\label{fig:ppc_m3_true}
\includegraphics[width=\textwidth]{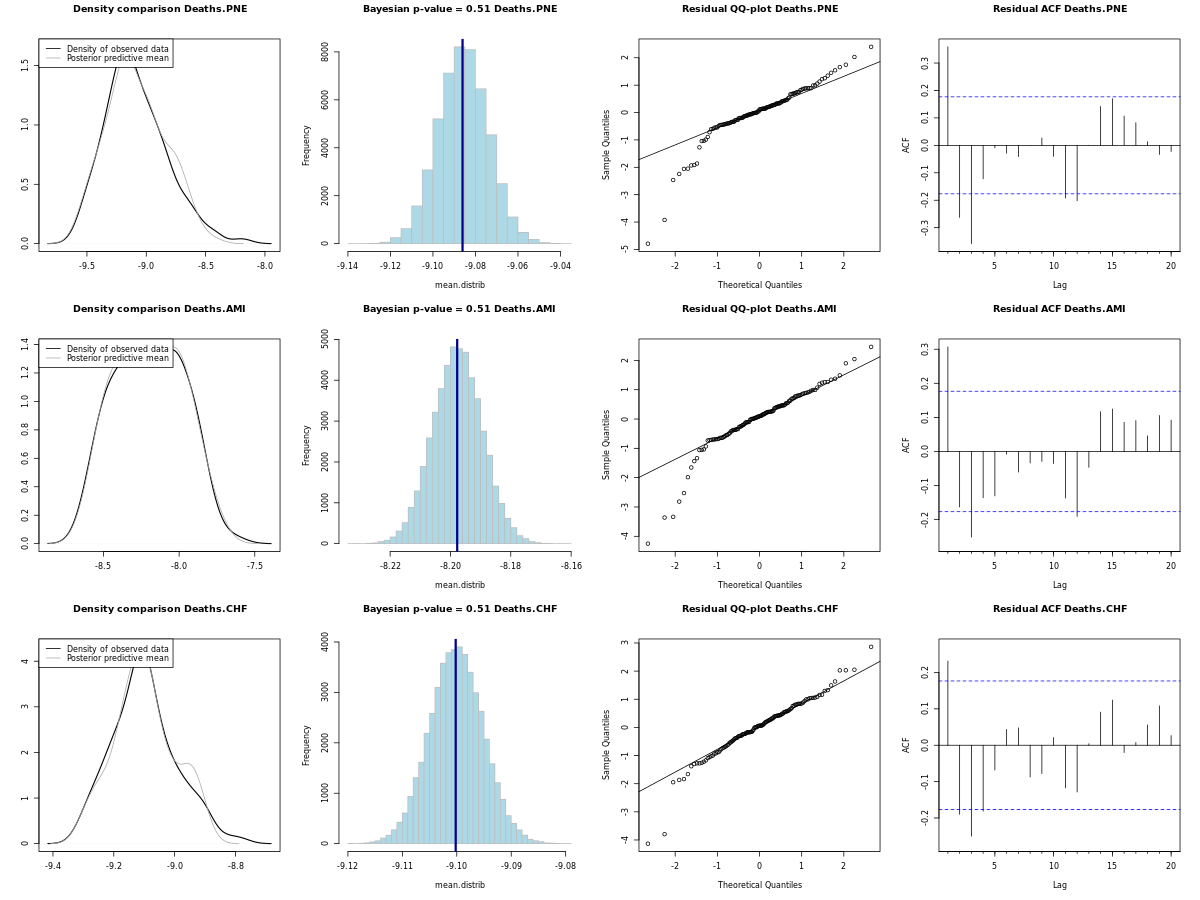}
\end{figure}

\begin{figure}[p]
\centering
\caption{Posterior predictive checks of control outcomes under Model 3.}
\label{fig:ppc_m3}
\includegraphics[width=\textwidth]{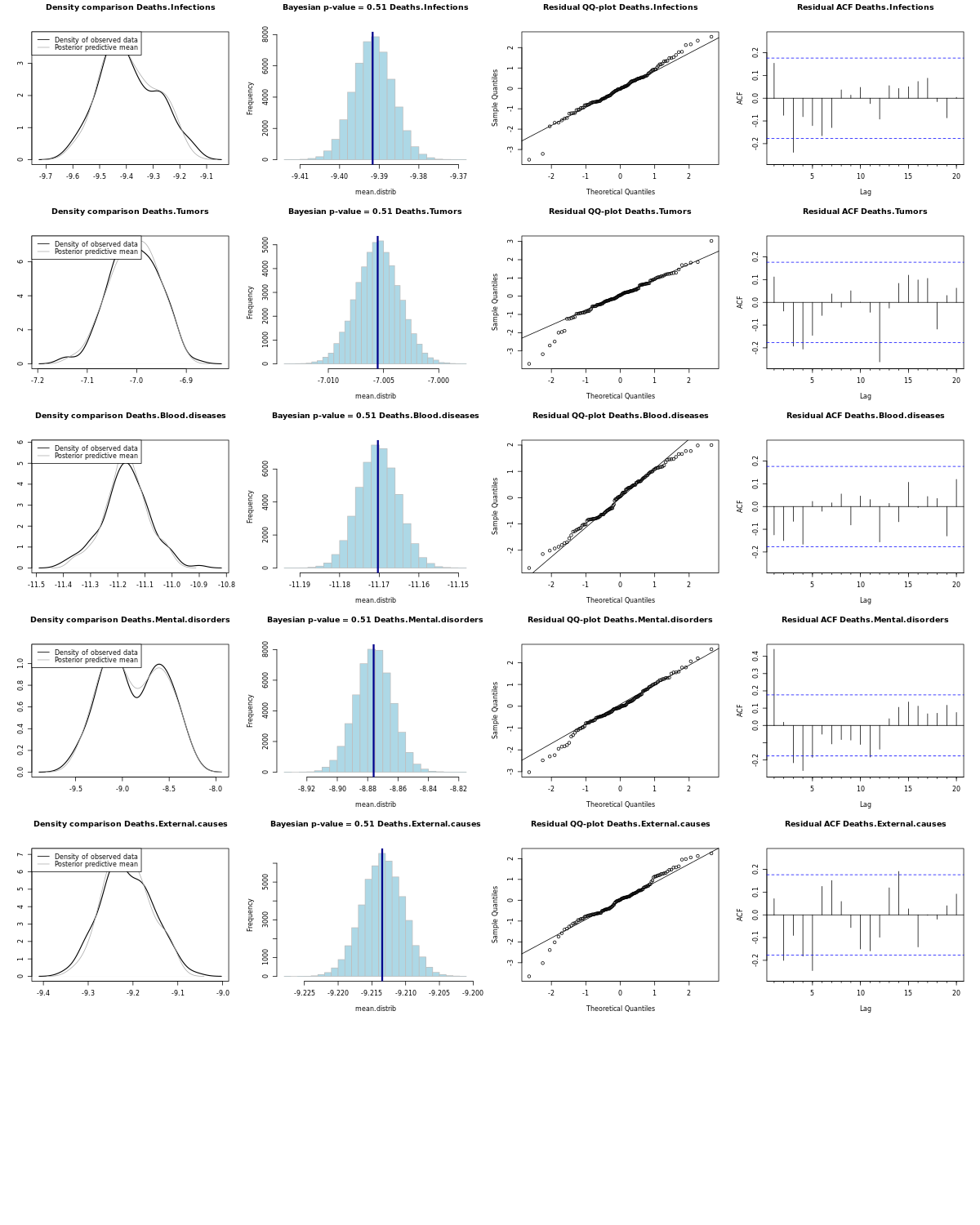}
\end{figure}

\begin{figure}
\centering
\caption{Trace plots of control outcomes, Model 1.}
\label{fig:trace_m1}
\includegraphics[width=\textwidth]{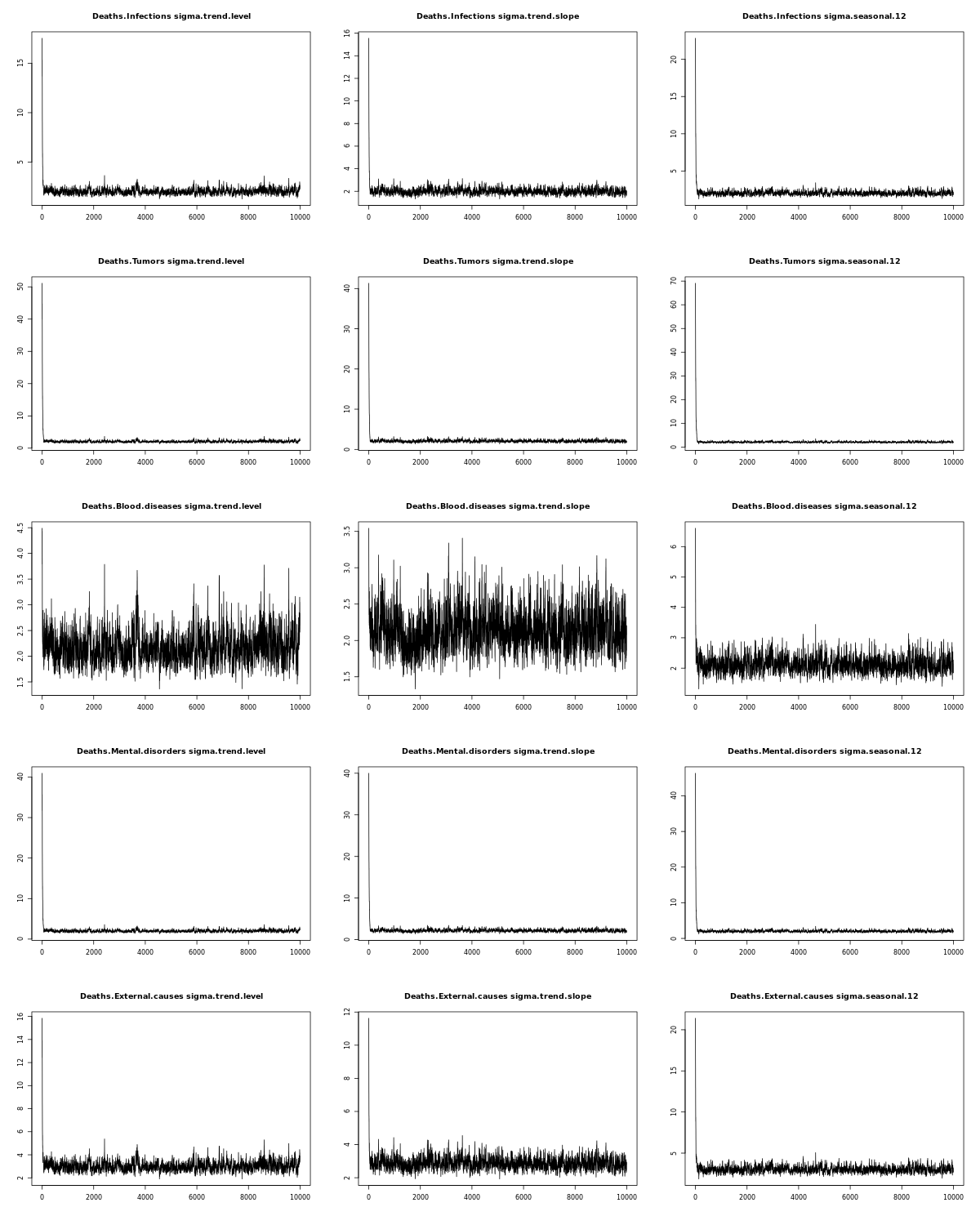}
\end{figure}

\begin{figure}
\centering
\caption{Trace plots of pneumonia, AMI and CHF, Model 2.}
\label{fig:trace_m2_true}
\includegraphics[width=\textwidth]{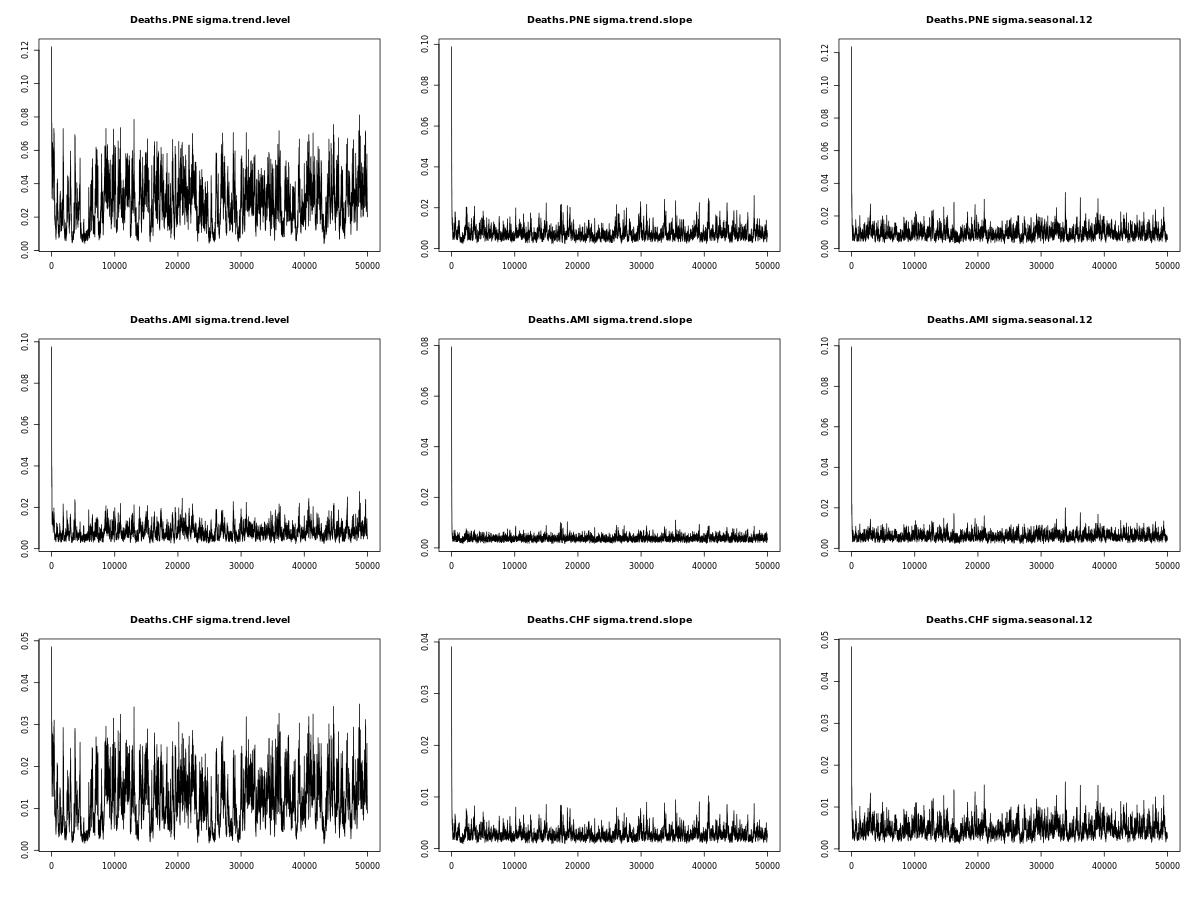}
\end{figure}

\begin{figure}
\centering
\caption{Trace plots of control outcomes, Model 2.}
\label{fig:trace_m2}
\includegraphics[width=\textwidth]{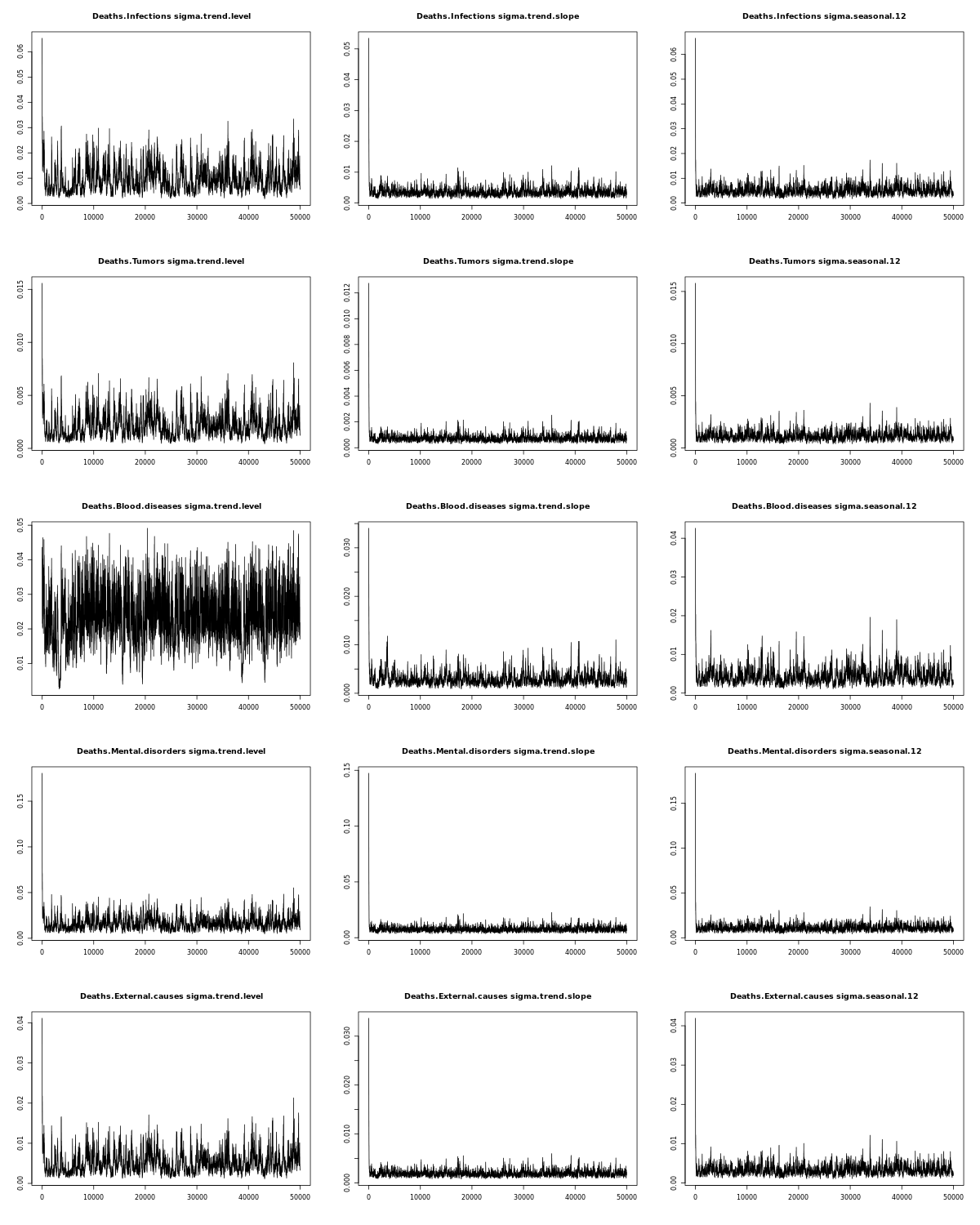}
\end{figure}

\begin{figure}
\centering
\caption{Trace plots of pneumonia, AMI and CHF, Model 3.}
\label{fig:trace_m3_true}
\includegraphics[width=\textwidth]{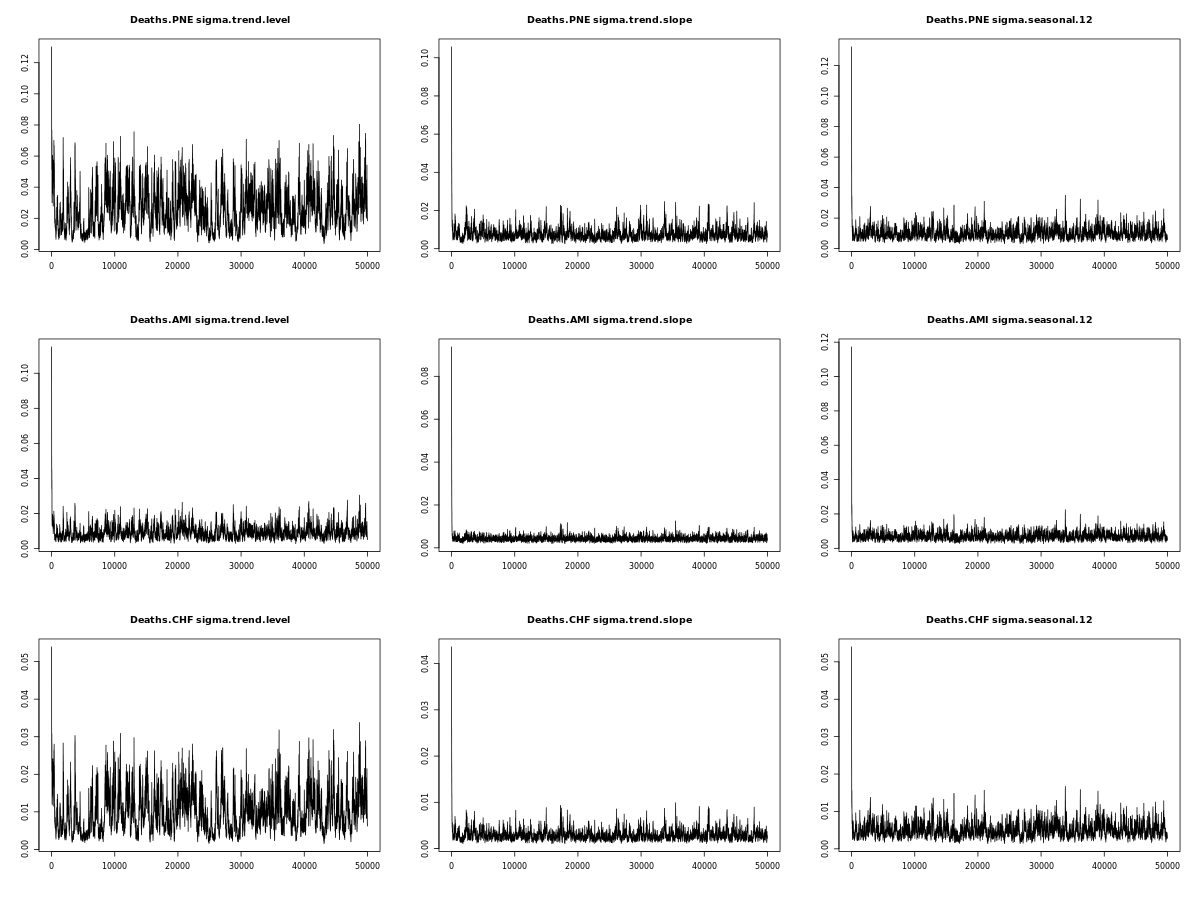}
\end{figure}

\begin{figure}
\centering
\caption{Trace plots of control outcomes, Model 3.}
\label{fig:trace_m3}
\includegraphics[width=\textwidth]{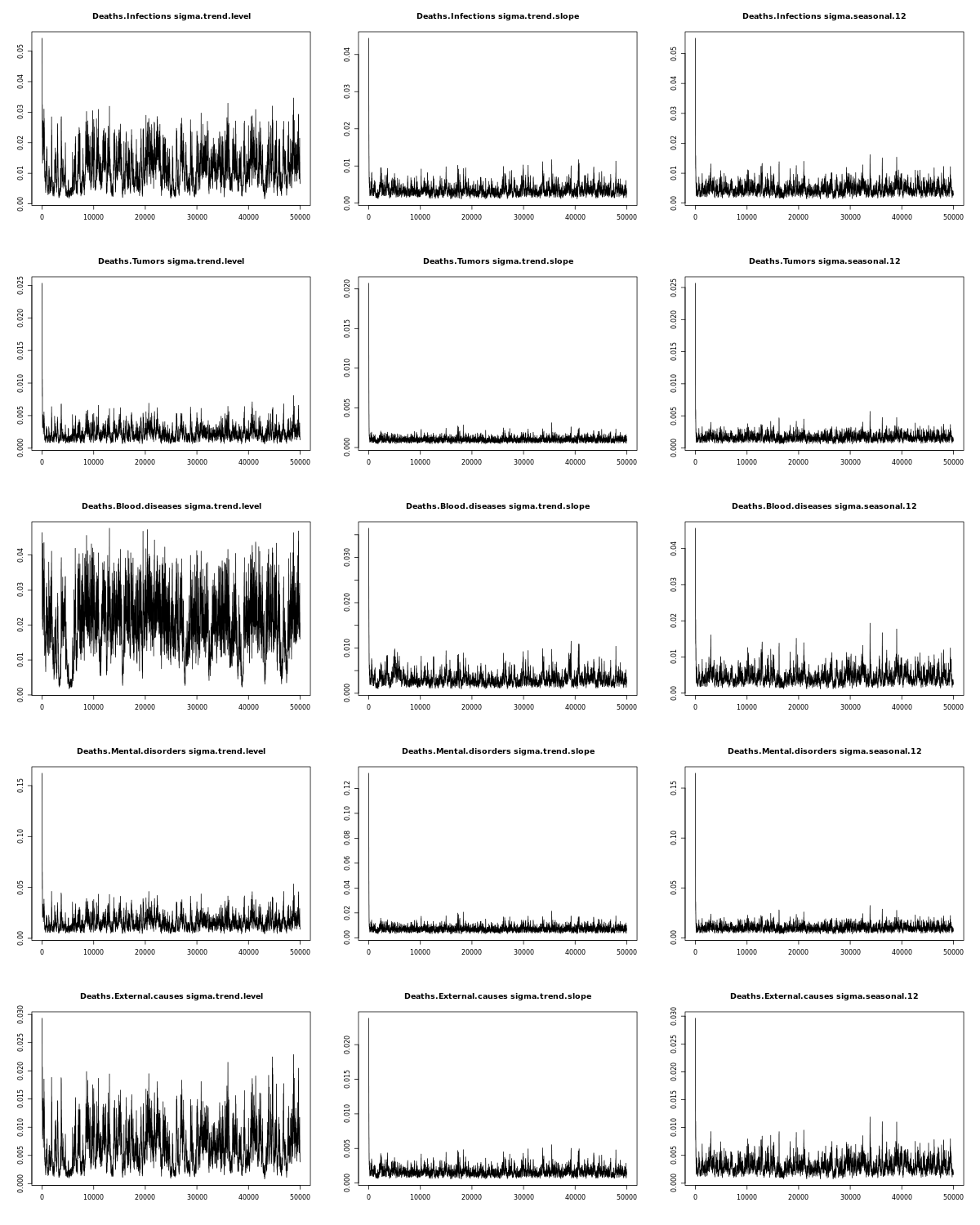}
\end{figure}

All the tested models include covariates. The posterior predictive checks and MCMC diagnostics for Model 1 on the target conditions are inspected in Sections \ref{subsec:ppc} and \ref{subsec:convergence}, respectively. \cref{fig:ppc_m1} shows the posterior predictive checks for Model 1 on the control conditions, Figures \ref{fig:ppc_m2_true} and \ref{fig:ppc_m2} show the posterior predictive checks for Model 2 for the target and control conditions respectively, and Figures \ref{fig:ppc_m3_true} and \ref{fig:ppc_m3} show the same quantities based on Model 3.
Figures \ref{fig:trace_m1}--\ref{fig:trace_m3} show the MCMC convergence checks for the models in the same order.
We see that all models converged and seem to provide a good fit to the pre-intervention data. Residuals diagnostics are also good, with a slightly better Normal approximation for Model 1, whereas the worst approximation seems to be under Model 3 (see for example the Normal QQ plots for the external causes and blood diseases under Model 3 compared to Model 1). This might be explained by the different frequency for the numerator (monthly death counts) and the denominator (yearly population estimates) that are used to compute mortality rates.

\clearpage

\clearpage

The results from Model 1 are reported in the main paper: Table \ref{tab:res} shows the results for pneumonia, AMI and CHF, and Table \ref{tab:res_falsification} for the control outcomes. The results from Models 2 and 3 are shown, respectively, in Tables \ref{tab:res_falsification_log} and \ref{tab:res_falsification_logit}.

\begin{table}[!b]
\centering
\caption{Falsification tests. Cumulative causal effect under Model 2 of the HRRP on pneumonia, AMI, CHF and control outcomes at three different time horizons.}
\label{tab:res_falsification_log}
\begin{adjustbox}{width=1\textwidth}
\begin{tabular}{lrrrrrrrrrrr}
  \hline
  & \multicolumn{3}{c}{Short term} & & \multicolumn{3}{c}{Mid term} & & \multicolumn{3}{c}{Long term} \\ \cline{2-4} \cline{6-8} \cline{10-12}
 & C$\Delta$ & 2.5\% & 97.5\% & & C$\Delta$ & 2.5\% & 97.5\% & & C$\Delta$ & 2.5\% & 97.5\% \\ 
  \hline
Infections       & 3.34      & -2.52     & 9.47  & & 24.57     & -55.67    & 109.83 & & 61.96     & -214.54   & 347.36 \\
Tumors           & -0.02     & -1.37     & 1.28  & & -0.04     & -17.23    & 17.13  & & -0.75     & -59.68    & 56.74  \\
Blood diseases   & 2.08      & -4.19     & 8.26  & & 12.55     & -61.85    & 86.91  & & 30.27     & -218.14   & 279.57 \\
Mental disorders & 2.55      & -9.90     & 15.19 & & 8.26      & -161.50   & 180.53 & & -2.75     & -586.00   & 583.10 \\
External causes  & 1.56      & -1.79     & 4.90  & & 10.72     & -33.35    & 55.35  & & 28.37     & -122.29   & 179.93 \\
PNE              & 4.82      & -8.19     & 18.30 & & 31.00     & -143.80   & 217.25 & & 71.56     & -524.20   & 689.68 \\
AMI              & 1.58      & -4.80     & 7.93  & & 12.92     & -71.28    & 97.64  & & 35.55     & -255.12   & 325.17 \\
CHF              & 0.53      & -4.85     & 5.65  & & 5.80      & -62.20    & 73.23  & & 12.99     & -220.14   & 239.62 \\
   \hline
\end{tabular}
\end{adjustbox}
\end{table}

\begin{table}[!b]
\centering
\caption{Falsification tests. Cumulative causal effect under Model 3 of the HRRP on pneumonia, AMI, CHF and on control outcomes at three different time horizons.}
\label{tab:res_falsification_logit}
\begin{adjustbox}{width=1\textwidth}
\begin{tabular}{lrrrrrrrrrrr}
  \hline
  & \multicolumn{3}{c}{Short term} & & \multicolumn{3}{c}{Mid term} & & \multicolumn{3}{c}{Long term} \\ \cline{2-4} \cline{6-8} \cline{10-12}
 & C$\Delta$ & 2.5\% & 97.5\% & & C$\Delta$ & 2.5\% & 97.5\% & & C$\Delta$ & 2.5\% & 97.5\% \\ 
  \hline
Infections       & 22.60     & -57.12    & 108.52 & & 22.60     & -57.12    & 108.52 & & 56.28     & -214.23   & 341.62 \\
Tumors           & -1.06     & -23.67    & 21.92  & & -1.06     & -23.67    & 21.92  & & -4.29     & -82.59    & 73.78  \\
Blood diseases   & 11.63     & -65.20    & 88.87  & & 11.63     & -65.20    & 88.87  & & 27.14     & -238.31   & 290.89 \\
Mental disorders & 6.89      & -151.77   & 169.76 & & 6.89      & -151.77   & 169.76 & & -2.61     & -551.91   & 549.90 \\
External causes  & 8.74      & -29.90    & 48.35  & & 8.74      & -29.90    & 48.35  & & 22.41     & -109.14   & 156.49 \\
PNE              & 27.35     & -148.24   & 212.29 & & 27.35     & -148.24   & 212.29 & & 68.96     & -534.68   & 694.65 \\
AMI              & 11.22     & -84.77    & 108.43 & & 11.22     & -84.77    & 108.43 & & 32.57     & -301.70   & 364.83 \\
CHF              & 4.06      & -65.13    & 73.23  & & 4.06      & -65.13    & 73.23  & & 9.25      & -227.97   & 242.20 \\
   \hline
\end{tabular}
\end{adjustbox}
\end{table}

At first sight, it might seem strange that Model 1 evidences some effect of the HRRP on pneumonia and AMI while Model 2 and 3 do not seem to spot any impact. However, this difference can be explained once we carefully consider the quantities that Models 1--3 estimate. Indeed, Model 1 targets an effect at the additive scale, while Models 2-3 target an effect at the multiplicative scale. More formally, let $Y_t$ be the death counts from a generic disease at time $t$ and let $y_t = \log{Y_t}$. In addition, denote with $R_t$ the mortality rate and let $r_t = \logit R_t$ denote its logit transformation, i.e.,
$$\logit R_t = \log{\frac{R_t}{1-R_t}}.$$
Then, under Model 1 the causal effect that we are estimating is:
$$\hat{\tau}_t^{(1)} = Y_t(1) - Y_t(0),$$
under Model 2 the effect is:
$$\hat{\tau}_t^{(2)} = y_t(1) - y_t(0) = \log{\frac{Y_t(1)}{Y_t(0)}},$$
and under Model 3 it is:
$$\hat{\tau}_t^{(3)} = r_t(1) - r_t(0)  = \log{ \frac{R_t(1)/(1-R_t(1))}{R_t(0)/(1-R_t(0))}} = \log{\Omega_t}$$
where $\Omega_t$ is the odds ratio (i.e., the ratio between the odds of dying and the odds of dying that over 65 people would have if HRRP was never adopted). Therefore, our results seem to show that there is no multiplicative effect of the treatment and there is no indication of an increase/decrease in the odds ratio, meaning that the odds of dying wouldn't have been higher/lower in the absence of the HRRP. However, the data support the conclusion that there is a positive additive effect of the death counts.

\end{document}